\documentclass[%
 aip,
 amsmath,amssymb,
 reprint,%
]{revtex4-1}

\usepackage{graphicx}% Include figure files
\usepackage{dcolumn}% Align table columns on decimal point
\usepackage{bm}% bold math
\usepackage[utf8]{inputenc}
\usepackage[T1]{fontenc}
\usepackage{mathptmx}
\usepackage{etoolbox}
\usepackage{multirow}
\usepackage{braket}
\usepackage{hyperref}
\usepackage{cleveref}
\makeatletter
\def\@email#1#2{%
 \endgroup
 \patchcmd{\titleblock@produce}
  {\frontmatter@RRAPformat}
  {\frontmatter@RRAPformat{\produce@RRAP{*#1\href{mailto:#2}{#2}}}\frontmatter@RRAPformat}
  {}{}
}%
\makeatother
\begin{document}

\title{Transcorrelated Methods Applied to Second Row Elements}
\author{Maria-Andreea Filip}
\email{a.filip@fkf.mpg.de}
\author{Pablo López Ríos}
\author{J. Philip Haupt}
\author{Evelin Martine Corvid Christlmaier}
\author{Daniel Kats}
\affiliation{Max Planck Institute for Solid State Research, Heisenbergstr. 1, 70569 Stuttgart, Germany}
\author{Ali Alavi}
\email{a.alavi@fkf.mpg.de}
\affiliation{Max Planck Institute for Solid State Research, Heisenbergstr. 1, 70569 Stuttgart, Germany}
\affiliation{Yusuf Hamied Department of Chemistry, University of Cambridge, Lensfield Road, Cambridge CB2 1EW, United Kingdom}

\date{\today}% It is always \today, today,

\begin{abstract}
     We explore the applicability of the transcorrelated method to the elements in the second row of the periodic table. We use transcorrelated Hamiltonians in conjunction with full configuration interaction quantum Monte Carlo and coupled cluster techniques to obtain total energies and ionisation potentials, investigating their dependence on the nature and size of the basis sets used. Transcorrelation accelerates convergence to the complete basis set limit relative to conventional approaches, and chemically accurate results can generally be obtained with the cc-pVTZ basis, even with a frozen Ne core in the post-Hartree--Fock treatment. 
\end{abstract}

\maketitle

\section{Introduction}
Among the many challenges still facing \textit{ab-initio} quantum chemistry methods is their slow convergence to the basis-set and full configuration interaction (FCI) limits. This is fundamentally due to the presence of singularities in the Coulomb potential whenever two charged particles coalesce, which leads to the so-called Kato cusp conditions,\cite{Kato1957} inducing discontinuities in the true wavefunction that are challenging to describe using Gaussian basis functions.

This problem can be mitigated by explicitly including a functional dependence on inter-particle distances in the wavefunction. This is done in the so-called explicitly correlated methods, the most well-known of which are the R12 and F12a approaches.\cite{Hylleraas1928, Hylleraas1929, Hylleraas1930,Kutzelnigg1985,Klopper1987,Klopper1990,Kutzelnigg1991,Noga1994,Noga1997,Ten-no2004,Valeev2004,Kedzuch2005,Adler2007,Knizia2008,Knizia2009,Hattig2010,Kong2012}

Another technique, known as transcorrelation (TC), was first introduced by Boys and Handy in the late 1960s.\cite{Boys1969,Handy1969,Handy1971} In this method, one uses a direct correlation function called a Jastrow factor\cite{Jastrow1955} to similarity transform the system Hamiltonian. The resulting operator is, however, non-hermitian, potentially leading to non-variational energies, and contains additional higher-order many-body terms that make its computation challenging. 

More recently, there has been a resurgence of interest in the TC approach.\cite{Nooijen1998,Tenno2000,Tenno2000b,Hino2002,Umezawa2003,Umezawa2004,Sakuma2006,Tsuneyuki2008, Luo2018,Cohen2019,Dobrautz2022,Ammar2022,Ammar2023,Ammar2023b,Lee2023, Ammar2024} This has led to the observation that the non-hermitian TC Hamiltonian has more compact right eigenvectors than the original Hamiltonian, which implies a formally decreased multi-configurational character of the wavefunction.\cite{Dobrautz2019} A variety of electronic structure methods can successfully treat this non-hermitian Hamiltonian, including Full Configuration Interaction Quantum Monte Carlo (FCIQMC),\cite{Cohen2019} Selected CI,\cite{Ammar2022,Ammar2023b,Ammar2024} Coupled Cluster (CC)\cite{Liao2021,Schraivogel2021,Schraivogel2023} and Density Matrix Renormalisation Group (DMRG)  algorithms.\cite{Baiardi2022,Liao2023} These approaches have been benchmarked on a variety of model systems,\cite{Luo2018,Dobrautz2019, Liao2021} first-row atoms\cite{Cohen2019,Schraivogel2021} and molecules,\cite{Gunther2021,Schraivogel2023, Liao2023, Haupt2023} showing improved quality results for a given correlation treatment, as well as accelerated convergence with basis set size.

With the introduction of flexible Jastrow factors, optimisable by Variational Monte Carlo (VMC),\cite{Haupt2023} and new approximations that allow for the efficient treatment of the three-body terms in the TC Hamiltonian,\cite{Christlmaier2023} it is now possible to extend this treatment to atoms beyond the first row. In this paper we report transcorrelated results for absolute energies and ionisation potentials of the second row atoms, investigating the accuracy of TC approaches relative to their non-TC counterparts and explicitly correlated F12a variants,\cite{Kedzuch2005,Adler2007,Werner2007,Knizia2009,Kats2015} where those are available. 
\begin{figure*}
    \centering
    \includegraphics[width=\textwidth]{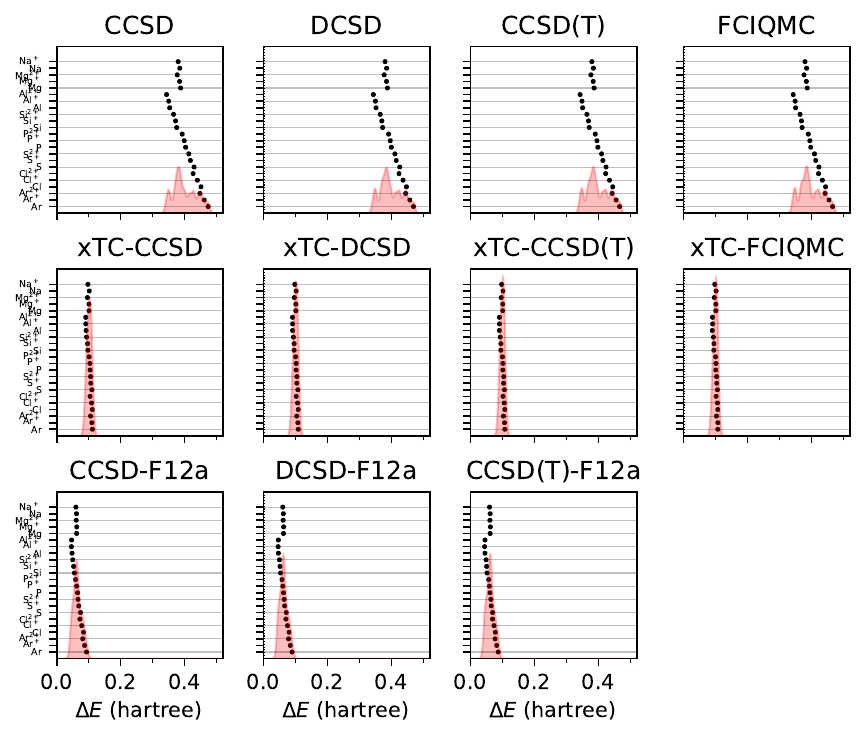}
    \caption{\small Error in the total energies of second-row atoms, as well as singly and doubly ionised species, relative to the values in Ref. \cite{Chakravoty1993}. Energies were computed using the cc-pVTZ basis set, with Jastrow cutoff terms $L_u = 4.5 a_0$, $L_\chi=2.5 a_0$ and $L_f = 4.0 a_0$. Black dots denote individual values, while the shaded area gives the resulting distribution.}
    \label{fig:tz_total}
\end{figure*}
\section{Theory}
\label{sec:theory}
\subsection{Transcorrelation}
Transcorrelated methods are based on the Slater-Jastrow form of the wavefunction,\cite{Jastrow1955}
\begin{equation}
    \Psi = e^J\Phi,
\end{equation}
where $\Phi$ is some anti-symmetrised wavefunction, commonly taken to be a single Slater determinant or a configuration interaction (CI) expansion and $J(\mathbf{r}_1,\mathbf{r}_2,...,\mathbf{r}_{N_e};\mathbf{R}_1,\mathbf{R}_2,...,\mathbf{R}_{N_a}) $ is a symmetric correlator depending on the positions of the $N_e$ electrons in the system and parametrically on the $N_a$ atomic positions. For simplicitly, we omit this dependence in the following discussion. 

This form of the wavefunction can be substituted into the Schr\"odinger equation,
\begin{equation}
    \hat H e^J \Phi = E e^J \Phi
\end{equation}
and pre-multiplying by $e^{-J}$ leads to the similarity transformed Schr\"odinger equation,
\begin{equation}
    \hat H_\text{TC} \Phi = E \Phi,
\end{equation}
where
\begin{equation}
   \hat H_\text{TC} = e^{-J} \hat H e^J.
\end{equation}
\label{sec:tc}
The transcorrelated Hamiltonian can be computed by taking a Baker–Campbell–Hausdorff expansion, which terminates after the second commutator, 
\begin{equation}
   \hat H_\text{TC} = \hat H + [\hat H, J] + \frac{1}{2} [[\hat H, J]J]. 
\end{equation}

Consider a Jastrow factor with up to two-electron terms, $J = \sum_{i<j}u(\mathbf{r}_i, \mathbf{r}_j)$, where $u(\mathbf{r}_i, \mathbf{r}_j)$ is some symmetric function. This leads to a Hamiltonian of the form
\begin{equation}
    \hat H_\text{TC} = \hat H - \sum_{i<j}\hat K(\mathbf{r}_i, \mathbf{r}_j) - \sum_{i<j<k} \hat L(\mathbf{r}_i, \mathbf{r}_j, \mathbf{r}_k),
\end{equation}
where 
\begin{align}
\begin{split}
   \hat K(\mathbf{r}_i, \mathbf{r}_j)= &\frac{1}{2}[\nabla_i^2u(\mathbf{r}_i, \mathbf{r}_j)+\nabla_j^2u(\mathbf{r}_i, \mathbf{r}_j)\\
    &+|\nabla_iu(\mathbf{r}_i, \mathbf{r}_j)|^2+|\nabla_ju(\mathbf{r}_i, \mathbf{r}_j)|^2]\\
    &+\nabla_iu(\mathbf{r}_i, \mathbf{r}_j)\cdot \nabla_i+\nabla_ju(\mathbf{r}_i, \mathbf{r}_j)\cdot \nabla_j
\end{split}
\end{align}
and
\begin{align}
    \begin{split}
        \hat L(\mathbf{r}_i, \mathbf{r}_j, \mathbf{r}_k)& = \nabla_iu(\mathbf{r}_i, \mathbf{r}_j)\cdot \nabla_iu(\mathbf{r}_i, \mathbf{r}_k)\\
        &+ \nabla_ju(\mathbf{r}_j, \mathbf{r}_k)\cdot \nabla_ju(\mathbf{r}_j, \mathbf{r}_i)\\
        &+ \nabla_ku(\mathbf{r}_k, \mathbf{r}_i)\cdot \nabla_ku(\mathbf{r}_k, \mathbf{r}_j)
    \end{split}
\end{align}
\begin{table*}
\caption{\small Mean absolute error (MAE), standard deviation (STD) and maximum error (ME) for total energies computed with various methods for the cc-pVTZ basis set.  Set A and B refer to two different sets of cutoffs used in the Jastrow factors, see text. All values are in mHartree.}
\label{tab:tz-errors}
\centering
    \begin{tabular}
{l|d|d|d|d|d|d|d|d|d}
     \multirow{2}{*}{Method} & \multicolumn{3}{|c|}{$E_\text{atom}$}&\multicolumn{3}{|c|}{$E_\text{monocation}$}&\multicolumn{3}{|c}{$E_\text{dication}$} \\
     \cline{2-10}
         & \mathrm{MAE} & \mathrm{STD} & \mathrm{MaxE} & \mathrm{MAE} & \mathrm{STD} & \mathrm{MaxE} & \mathrm{MAE} & \mathrm{STD} & \mathrm{MaxE}   \\
         \hline
     CCSD &407.3&40.8&473.6&400.6&37.1&461.4&395.2&36.4&447.8\\
     DCSD& 404.7&39.5&468.9&398.6&35.7&457.3&393.6&35.1&444.4\\
     CCSD(T)&403.5&38.9&466.6&397.8&35.3&455.7&393.0&34.9&443.2\\
     FCIQMC&403.0&38.7&466.1&397.4&35.0&455.0&392.6&34.6&442.6\\
     \hline
     CCSD-F12a&68.2&14.9&93.5&64.9&12.2&87.3&62.6&12.2&81.4\\
     DCSD-F12a&65.8&13.9&89.2&63.1&12.1&83.7&61.1&11.1&78.2\\
     CCSD(T)-F12a&64.5&13.4&87.0&62.2&11.9&82.1&60.4&11.0&77.1\\
     \hline
     (A) xTC-CCSD &117.2&10.4&133.3&113.1&9.2&127.4&109.2&9.2&121.9\\
     (A) xTC-DCSD &114.9&9.4&129.6&111.4&8.0&124.2&107.9&8.0&119.1\\
     (A) xTC-CCSD(T)  &114.2&9.0&127.9&111.0&7.7&123.1&107.6&7.9&118.4\\
     (A) xTC-FCIQMC  &113.6&8.9&127.4&110.5&7.4&122.4&107.2&7.5&117.6\\
      \hline
     (B) xTC-CCSD &103.8&7.0&111.7&102.0&6.6&110.0&99.8&6.2&106.0\\
     (B) xTC-DCSD &101.9&6.3&108.7&100.6&5.6&107.3&98.7&5.2&104.2\\
     (B) xTC-CCSD(T) &101.4&6.0&107.6&100.4&5.4&106.6&98.6&5.2&104.2\\
     (B) xTC-FCIQMC &101.1&6.1&107.1&100.4&5.6&107.4&98.4&5.0&103.4\\
     \hline
     \end{tabular}
\end{table*}
In second quantisation, for a set of one-particle basis functions $\{\phi_p\}$, the resulting Hamiltonian operator is given by
\begin{equation}
\begin{split}
    \hat H_\text{TC} &= \sum_{pq} h^p_{q}\hat a_p^\dagger \hat a_q + \frac{1}{2}\sum_{pqrs} (V^{pq}_{rs}-K^{pq}_{rs})\hat a_p^\dagger \hat a_q^\dagger \hat a_s \hat a_r \\
    &- \frac{1}{6} \sum_{pqrstu} L_{stu}^{pqr}\hat a_p^\dagger \hat a_q^\dagger \hat a_r^\dagger \hat a_u \hat a_t \hat a_s,
    \end{split}
\end{equation}
where $h^{p}_{q} = \braket{\phi_p|\nabla^2|\phi_q}$, $V^{pq}_{rs} = \braket{\phi_p\phi_q|r_{12}^{-1}|\phi_r\phi_s}$, ${K^{pq}_{rs} = \braket{\phi_p\phi_q|\hat K|\phi_r\phi_s}}$ and ${L^{pqr}_{stu} = \braket{\phi_p\phi_q\phi_r|\hat L|\phi_s\phi_t\phi_u}}$.
The presence of the additional three-body operator $\hat L$ makes the computation of the TC Hamiltonian significantly costlier than that of the underlying Hamiltonian. However, it has been shown that excluding the explicit three-body contributions has little effect on the resulting energies.\cite{Schraivogel2021,Schraivogel2023} Under the so-called xTC approximation\cite{Christlmaier2023} it is then possible to fold the remaining three-body terms into the lower-order integrals. When considering a generalised normal ordered form of the TC Hamiltonian with respect to some reference wavefuntion $\ket{\Phi}$,
\begin{equation}
    \hat H_\text{N} = \hat H_\text{TC} - \braket{\Phi|\hat H_\text{TC}|\Phi} = \hat F_\text{N} + \hat V_\text{N}+ \hat L_\text{N},
\end{equation}
the one-, two- and three- body operators can be written as
\begin{equation}
\begin{split}
    \hat F_\text{N} &= [h_q^p + (U_{qs}^{pr} - U_{sq}^{pr})\gamma_s^r \\
    &- \frac{1}{2}(L_{qsu}^{prt}- L_{qus}^{prt}-L_{usq}^{prt})\gamma_{su}^{rt}]\widetilde a_q^p,
\end{split}
    \label{eq:xtc-one-body}
\end{equation}
\begin{equation}
    \hat V_\text{N} = \frac{1}{2}[U_{qs}^{pr} - (L_{qsu}^{prt}- L_{qus}^{prt}-L_{usq}^{prt})\gamma_{u}^{t}]\widetilde a_{qs}^{pr}
    \label{eq:xtc-two-body}
\end{equation}
and
\begin{equation}
    \hat L_\text{N} = \frac{1}{6}L_{qsu}^{prt} \widetilde a_{qsu}^{prt},
    \label{eq:xtc-three-body}
\end{equation}
    where Einstein summation convention has been employed, $U_{qs}^{pr} = K_{qs}^{pr} + V_{qs}^{pr}$, $\widetilde a_{q...}^{p...}$ are normal ordered excitation operators and the density matrices $\gamma_{q...}^{p...} = \braket{\Phi|\hat a_{q...}^{p...}|\Phi}$. Ignoring the three-body term in \cref{eq:xtc-three-body} and only computing the corrections in \cref{eq:xtc-one-body,eq:xtc-two-body} and a zero-body correction given by $\braket{\Phi|\hat L|\Phi}$ --- all of which can be done in practice without explicit computation of $L$ (see Ref. \cite{Christlmaier2023}) --- decreases the overall cost of obtaining the TC Hamiltonian from $\mathcal{O}(N_\mathrm{orb}^6 N_\mathrm{grid})$ to $\mathcal{O}(N_\mathrm{orb}^4 N_\mathrm{grid})$ with the number of orbitals $N_\mathrm{orb}$ and grid points $N_\mathrm{grid}$ used to describe the system.\cite{Christlmaier2023}
    
\subsection{Jastrow factor}
\label{sec:jastrow}
A variety of different forms for the Jastrow factor have been proposed over the years. In this work, we use the same form as Refs. \cite{Drummond2004,LopezRios2012,Haupt2023}, in which the Jastrow is given by a sum of electron-electron, electron-nucleus and electron-electron-nucleus terms,
\begin{equation}
    J = \sum_{i<j}u(r_{ij}) + \sum_{i,I}\chi(r_{iI}) + \sum_{i<j,I}f(r_{ij}, r_{iI}, r_{jI})
\end{equation}
expressed as natural power expansions
\begin{equation}
    u(r_{ij}) = t(r_{ij},L_u)\sum_k a_k r_{ij}^k,
    \label{eq:u}
\end{equation}
\begin{equation}
    \chi(r_{iI}) = t(r_{iI},L_\chi)\sum_k b_k r_{iI}^k
    \label{eq:chi}
\end{equation}
and
\begin{equation}
   f(r_{ij}, r_{iI}, r_{jI}) = t(r_{iI},L_f)t(r_{jI},L_f)\sum_{klm} c_{klm} r_{ij}^kr_{iI}^lr_{jI}^m,
    \label{eq:f}
\end{equation}

where $\{a_k\}$, $\{b_k\}$, $\{c_{klm}\}$ are linear coefficients, ${t(r,L)=(1-r/L)^3\Theta(L-r)}$, $\Theta(x)$ is the Heaviside function and $L_u$, $L_\chi$, $L_f$ are some appropriate cutoff radii.

\begin{figure*}
    \centering
    \includegraphics[width=\textwidth]{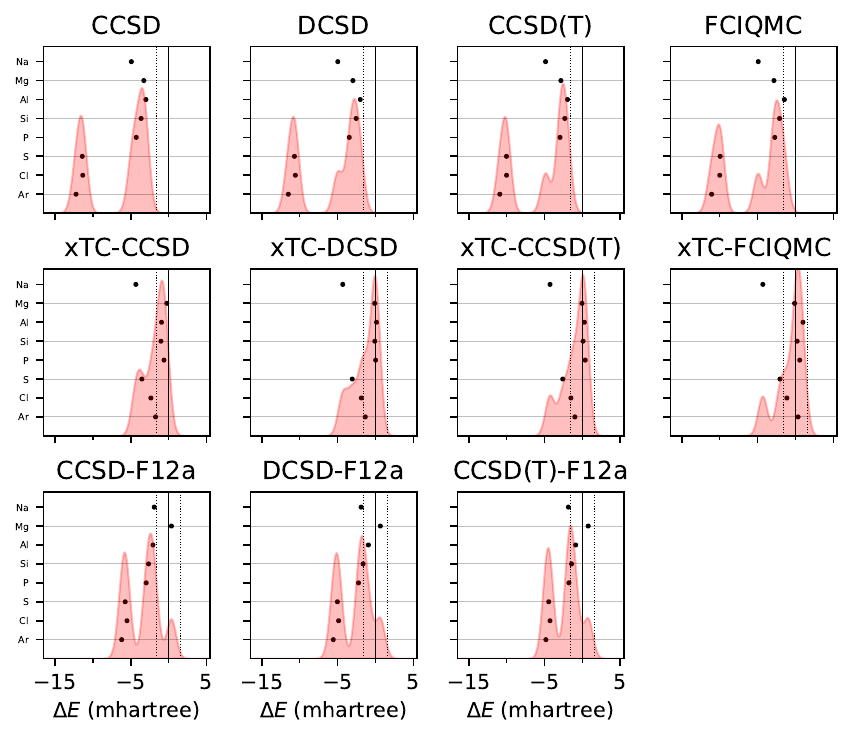}
    \caption{\small Error in the first ionisation energies of second-row atoms,  relative to values obtained from Ref. \cite{Chakravoty1993}. Energies were computed using the cc-pVTZ basis set, with Jastrow cutoff terms $L_u = 4.5 a_0$, $L_\chi=2.5 a_0$ and $L_f = 4.0 a_0$. Black dots denote individual values, while the shaded area gives the resulting distribution. The dotted lines denote chemical accuracy.}
    \label{fig:tz_ip1}
\end{figure*}
The electron-electron cusp condition is enforced through the $u$ term in the Jastrow factor. In principle, it is possible to enforce electron-nuclear cusps by fixing appropriate terms in the $\chi$ functions, but to obtain accurate variational energies it has been found to be better to either modify the $s$-like components of the molecular orbitals near nuclei to introduce a cusp\cite{Ma2005} or to include this transformation as an additional term in the Jastrow factor,\cite{Haupt2023} of the form
\begin{equation}
    \Lambda(r_{iI}) = [\ln{\widetilde \phi(r_{iI})}-\ln{\phi(r_{iI})}]\Theta(r_{iI}-r_c),
\end{equation}
where $\phi(r)$ is the $l = 0$ component of the molecular orbital, $r_c$ is a cutoff length and
\begin{equation}
    \widetilde \phi(r) = e^{\sum_{l=0}^4 \lambda_l r^l} + C,\ \ \ r< r_c.
\end{equation}
Details for how to obtain these parameters are given in Refs. \cite{Haupt2023, Ma2005}. 

The free parameters of the Jastrow factor are optimised using VMC. Rather than minimising the variational energy, 
\begin{equation}
    E = \frac{\braket{\Psi|\hat H | \Psi}}{\braket{\Psi|\Psi}},
\end{equation}
it is possible to minimise an alternative cost function, such as the variance. In this case, we optimise the so-called ``variance of the reference energy",\cite{Handy1971, Umezawa2003, Haupt2023}
\begin{equation}
    \sigma_\text{ref}^2 = \braket{\Phi_\text{HF}||(\hat H_\text{TC} - E_\text{ref})|^2|\Phi_\text{HF}},
\end{equation}
where $\hat H_\text{TC} = e^{-J}\hat H e^{J}$ and $E_\text{ref} = \braket{\Phi_\text{HF}|\hat H_\text{TC}|\Phi_\text{HF}}$. This has been shown to introduce lower stochastic noise in the optimisation than conventional variance minimisation.\cite{Haupt2023}
\begin{figure*}
    \centering
    \includegraphics[width=\textwidth]{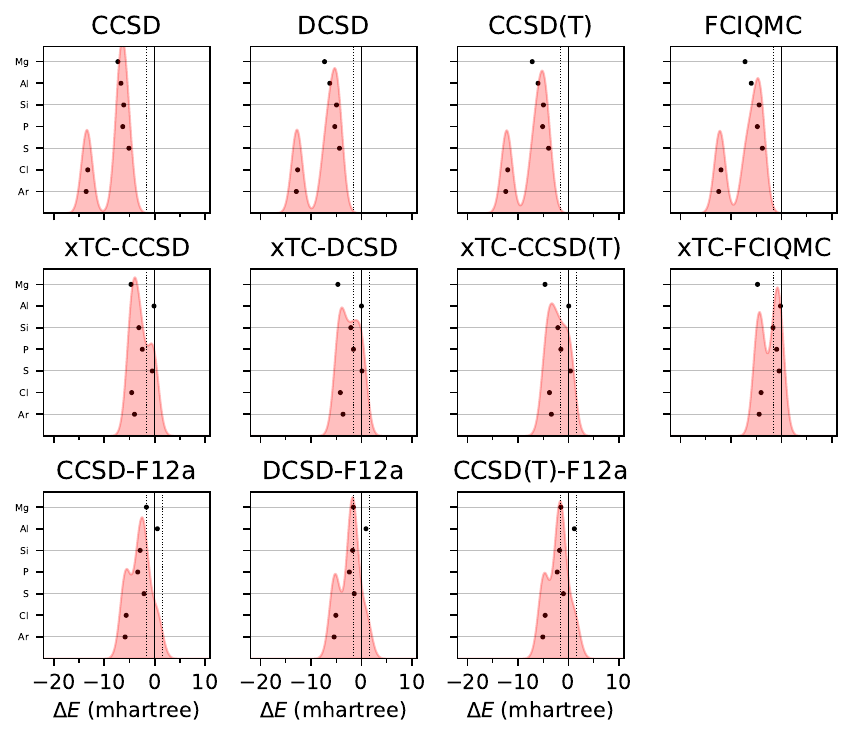}
    \caption{\small Error in the second ionisation energies of second-row atoms,  relative to values obtained from Ref. \cite{Chakravoty1993}. Energies were computed using the cc-pVTZ basis set, with Jastrow cutoff terms $L_u = 4.5 a_0$, $L_\chi=2.5 a_0$ and $L_f = 4.0 a_0$. Black dots denote individual values, while the shaded area gives the resulting distribution. The dotted lines denote chemical accuracy.}
    \label{fig:tz_ip2}
\end{figure*}

\section{Results and Discussion}
\label{sec:results}

\subsection{Computational details}
We compute the absolute energies of the second row atoms, with atomic number $Z = 11-18$, as well as their corresponding singly and doubly ionised species using the transcorrelated approach, under the xTC approximation. The Jastrow factor contains 46 free linear parameters, and is optimized in the presence of the canonical Hartree-Fock
determinant by VMC-based minimisation of the variance of the
reference energy. The cutoff lengths in \Cref{eq:u,eq:chi,eq:f} were set to $L_u = 4.5 a_0$ and $L_\chi = L_f = 4.0 a_0$ in Ref.\ \cite{Haupt2023} for first row species. However, in the case of the second row and particularly for more highly charged species, we have found that these relatively large cutoffs may lead to instabilities in the subsequent correlation treatment. We present here results obtained with a conservative choice of $L_u = L_\chi = L_f = 2.0 a_0$ (set A). For such low cutoff values, correlation methods converge reliably, but results are noticeably less accurate than those obtained with large cutoff values, where such a calculation is stable. Further investigation shows that results are most sensitive to $L_\chi$, so we present further results with $L_u = 4.5 a_0$, $L_\chi=2.5 a_0$ and $L_f = 4.0 a_0$ (set B), which provide stable, high-accuracy results. Correlation effects beyond the Slater-Jastrow wavefunction are included using initiator FCIQMC\cite{Cleland2010} or coupled cluster methods. The resulting total xTC energies are given in the supplementary material.

The Na$^{2+}$ ion [$1s^22s^22p^5$] is the only species considered which has an incomplete 2p shell. As the cc-pVXZ (X = D, T, Q) basis sets considered in this paper are designed to account for valence correlation effects, the treatment of these formally core orbitals is sub-optimal. We therefore exclude this species from all further discussion, although we note that employing a core-valence basis set such as cc-pCVTZ\cite{Woon1993,Peterson2002,Prascher2011} can lead to accurate treatment of this species with TC approaches. In contrast, conventional methods go from overestimating the second Na IP to underestimating it in the presence of additional basis functions with core character.

Initial Hartree--Fock calculations were carried out using the PySCF package\cite{Sun2015,Sun2018,Sun2020} and the Jastrow parameters were optimised using CASINO.\cite{Needs2020} Transcorrelated integrals were obtained using the TCHINT library.\cite{tchint} FCIQMC calculations were carried out using the NECI package,\cite{Kai2020} F12 calculations using Molpro,\cite{MOLPRO-WIREs,MolproJCP,MOLPRO}
and coupled-cluster calculations were done using the ElemCo.jl package.\cite{elemcoil}

Absolute energies, as well as first and second ionisation potentials (IPs) are benchmarked against the highly accurate fully extrapolated estimates of Chakravoty \textit{et al}.\cite{Chakravoty1993}

\subsection{Comparison to other methods}

We begin by presenting error metrics for total (\cref{tab:tz-errors}) energies, as well as first and second (\cref{tab:tz-ip-errors}) ionisation potentials for the Na-Ar atoms, computed with conventional CCSD and CCSD(T), explicitly correlated CCSD-F12a, DCSD-F12a and CCSD(T)-F12a,  as well as xTC-CCSD,\cite{Schraivogel2021,Christlmaier2023} xTC-CCSD(T) (in the form of pcBO-CCSD(T) from Ref.~\cite{katsOrbital2024}), xTC-DCSD\cite{kats_dc_2013,kats_dcsd_2014,Schraivogel2021,Christlmaier2023} and xTC-FCIQMC in the cc-pVTZ basis set.\cite{Woon1993,Prascher2011} The distribution of set B errors in the total energies, first IP and second IP are given in Figures 1, 2 and 3, respectively. Underlying data for individual elements is given in the Supporting Information.

For total energies, both transcorrelated methods provide an error reduction of approximately a factor of 4 over conventional approaches, while also lowering the spread of energy errors. In this case, F12 methods are more effective in lowering the absolute energy errors, but provide less consistent results, with a consistently higher standard deviation (STD) than the comparable TC methods. This is different from behaviour observed for first-row molecules, where xTC energy errors were generally lower than F12 ones.\cite{Christlmaier2023} For the ionisation energies, error cancellation significantly reduces this discrepancy in result quality. For the two IPs, errors in the xTC (A) energies are approximately $1.5\times$ smaller than their conventional counterparts, while xTC (B) errors are about $4\times$ smaller, approaching chemical accuracy (1.6 mHartree). 

\begin{table}
\caption{\small Mean absolute error (MAE), root mean squared error (RMSD) and maximum error (ME) for ionisation energies of Na--Ar species computed with various methods for the cc-pVTZ basis set. The second IP of Na is excluded.  All values are in mHartree.}
\label{tab:tz-ip-errors}
\centering
    \begin{tabular}
{l|c|c|c|c|c|c}
     \multirow{2}{*}{Method} & \multicolumn{3}{|c|}{$E_{\text{IP}_1}$}&\multicolumn{3}{|c}{$E_{\text{IP}_2}$} \\
     \cline{2-7}
         & MAE & RMSD & MaxE & MAE & RMSD & MaxE   \\
         \hline
     CCSD &6.8&7.8&12.2&8.4&9.0&13.6\\
     DCSD &6.1&7.2&11.5&7.7&8.4&12.9\\
     CCSD(T)&5.7&6.8&10.9&7.7&8.3&12.4\\
     FCIQMC &5.6&6.8&11.1&7.3&8.0&12.4\\
     \hline
     CCSD-F12a&3.4&3.9&6.2&3.2&3.7&5.9\\
     DCSD-F12a&2.8&3.4&5.6&2.7&3.2&5.4\\
     CCSD(T)-F12a&2.5&3.0&4.8&2.4&3.0&5.1\\
      \hline
     (A) xTC-CCSD  &4.1&4.4&6.3&4.5&4.7&6.4\\
     (A) xTC-DCSD  &3.5&3.9&5.7&3.9&4.3&6.4\\
     (A) xTC-CCSD(T)  &2.6&3.0&5.2&3.7&4.0&6.3\\
     (A) xTC-FCIQMC  &3.1&3.6&5.2&3.5&3.9&6.4\\
      \hline
     (B) xTC-CCSD  &1.8&2.3&4.3&2.8&3.3&4.7\\
     (B) xTC-DCSD  &1.4&2.0&4.3&2.3&2.9&4.7\\
     (B) xTC-CCSD(T)  &1.3&1.8&4.3&2.3&2.8&4.6\\
     (B) xTC-FCIQMC  &1.2&1.8&4.3&2.3&3.0&4.8\\
     \hline
     \end{tabular}
\end{table}

The accuracy of F12 methods fall between those of the (A) and (B) xTC approaches, highlighting the value of mid-range Jastrow correlations. For all methods, errors are lowest for the early $p$-block elements (Al, Si, P), with all xTC (B) methods, as well as xTC-DCSD (A), xTC-FCIQMC (A), DCSD-F12 and CCSD(T)-F12 achieving chemical accuracy in these cases, and significantly higher for the heavier atoms (S, Cl, Ar). For the first IP, xTC (B) methods significantly lower these errors with respect to F12, but the second IP remains challenging for both approaches. 

A previous xTC study of molecules composed of first-row atoms\cite{Christlmaier2023}  found that xTC-DCSD significantly outperformed xTC-CCSD. While xTC-DCSD also yields better results in this case, the difference is much smaller than for first-row species both in
absolute and relative terms. The inclusion of the perturbative triples correction has a significant effect in the absolute energies, but it is reduced in the relative quantities. Finally, we note that all transcorrelated coupled cluster methods' errors are within 1 mHartree of the xTC-FCIQMC results, suggesting that the most important correlation effects are included at the double excitation level. 

\begin{table}[h]
\caption{\small Mean absolute error (MAE), root mean squared error (RMSD) and maximum error (ME) for ionisation energies computed with various methods for the cc-pVQZ basis set. All values are in mHartree.  Methods labelled ``FC" correspond to results with the Ne core frozen in the coupled cluster calculation. The first IP of Na and the second IP of Mg are excluded, as they are not well treated by FC methods.}
\label{tab:qz-ip-errors}
\centering

    \begin{tabular}
{l|c|c|c|c|c|c}

     \multirow{2}{*}{Method} & \multicolumn{3}{|c|}{$E_{\text{IP}_1}$}&\multicolumn{3}{|c}{$E_{\text{IP}_2}$} \\
     \cline{2-7}
         & MAE & RMSD & ME &MAE & RMSD & ME   \\
         \hline
         CCSD&4.5&4.8&6.7&5.7&5.8&7.8\\
         DCSD&3.6&3.9&5.8&5.1&5.4&7.9\\
         CCSD(T)&2.9&3.3&4.8&4.4&4.6&6.2\\
         \hline
      FC-CCSD&4.5&4.8&6.6&5.9&6.1&7.8\\
      FC-DCSD&3.6&4.0&5.7&5.1&5.4&7.1\\
      FC-CCSD(T)&3.0&3.4&4.7&4.6&5.0&6.7\\
      \hline
     CCSD-F12a&2.2&2.5&3.4&2.5&2.8&3.7\\
     DCSD-F12a&1.4&1.6&2.6&1.9&2.1&3.1\\
     CCSD(T)-F12a&0.8&0.9&1.6&1.5&1.6&2.3\\
     \hline
     FC-CCSD-F12a&2.9&3.0&3.9&4.4&4.5&5.9\\
     FC-DCSD-F12a&2.1&2.3&3.7&3.6&3.9&5.9\\
     FC-CCSD(T)-F12a&1.4&1.8&3.7&3.2&3.5&5.9\\
     \hline
     (A) xTC-CCSD&1.9&2.0&2.8&1.7&1.9&2.4\\
     (A) xTC-DCSD&1.2&1.3&2.1&1.0&1.2&2.0\\
     (A) xTC-CCSD(T)&0.6&0.8&1.5&1.0&1.1&1.9\\
     %(A) xTC-FCIQMC&0.6&0.8&1.4&0.8&0.9&1.8\\
     \hline
     (A) xTC-FC-CCSD&2.1&2.2&3.4&2.2&2.0&4.2\\
     (A) xTC-FC-DCSD&1.3&1.6&3.4&1.6&2.4&4.2\\
     (A) xTC-FC-CCSD(T)&0.9&1.4&3.3&1.4&1.9&4.1\\
     \hline
     (B) xTC-CCSD&0.9&1.0&2.0&1.3&1.6&2.3\\
     (B) xTC-DCSD&0.3&0.5&1.4&1.0&1.2&1.9\\
     (B) xTC-CCSD(T)&0.4&0.5&0.7&0.9&1.0&1.5\\
     \hline
     (B) xTC-FC-CCSD&0.7&0.8&1.4&1.3&1.5&2.3\\
     (B) xTC-FC-DCSD&0.6&0.6&0.9&1.0&1.0&1.3\\
     (B) xTC-FC-CCSD(T)&0.8&1.4&3.3&1.9&2.4&4.0\\
     
     \hline
     \end{tabular}
\end{table}

While the TC approach or explicit correlation significantly improves the performance of methods in the TZ basis set, it is nevertheless insufficient to obtain entirely chemically accurate results using the (A) Jastrow cutoff values. To achieve this, one must go to the cc-pVQZ basis set, results for which are shown in \cref{tab:qz-ip-errors}. In this case, all methods considered except xTC-CCSD (A) and CCSD-F12a show average errors below 1.6 mHartree, with maximum deviations of less than 2 mHartree. In this case, the difference between xTC (A) and F12 narrows, with xTC (A) methods actually out-performing the corresponding F12 methods. xTC-CCSD and xTC-DCSD (B) provide a further improvement, but the perturbative corrections in xTC-CCSD(T) (B) now lead to significant worsening of the results.

\subsection{Basis set size dependence}
\begin{figure}[h!]
    \centering
    \includegraphics[width=0.5\textwidth]{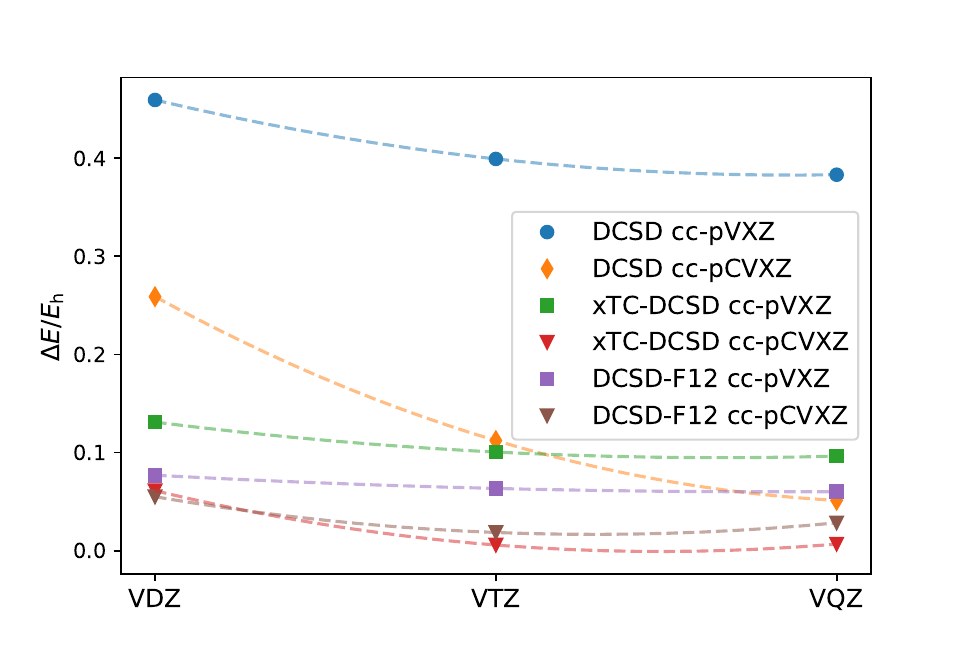}
    \caption{Basis set dependence of the mean absolute error in total energies $\Delta E$ computed with DCSD, DCSD-F12a and xTC-DCSD (B), relative to the values reported in Ref. \cite{Chakravoty1993}.}
    \label{fig:total_bde}
\end{figure}

The dependence of the mean absolute error (MAE) of the total energies on basis set is given in \cref{fig:total_bde} for the xTC (B). As was observed in the cc-pVTZ case above, differences between the methods employed are relatively minor on this scale, so we only show conventional CCSD and xTC-CCSD. The same trends hold for the other studied methods. In all cases, error decreases with basis set size. For the cc-pVXZ (X = D, T, Q) family, both conventional and xTC approaches seem to converge to total energies far from the exact values, although the error is significantly decreased by using a transcorrelated approach. Employing core-valence basis sets significantly improves this. In all basis sets, the transcorrelated approach leads to an approximately 4-fold reduction in the error. The convergence to the complete basis set limit is therefore accelerated, in particular for the core-valence basis sets. F12 methods exhibit similar trends and, as observed for the cc-pVTZ basis in \cref{tab:tz-errors}, generate more accurate energies in the cc-pVXZ basis family. However, when introducing additional core functions, the difference between the two methods all but disappears, with xTC giving better results for the largest cc-pCVQZ basis set.

\begin{figure*}
    \centering
    \includegraphics[width=0.8\textwidth]{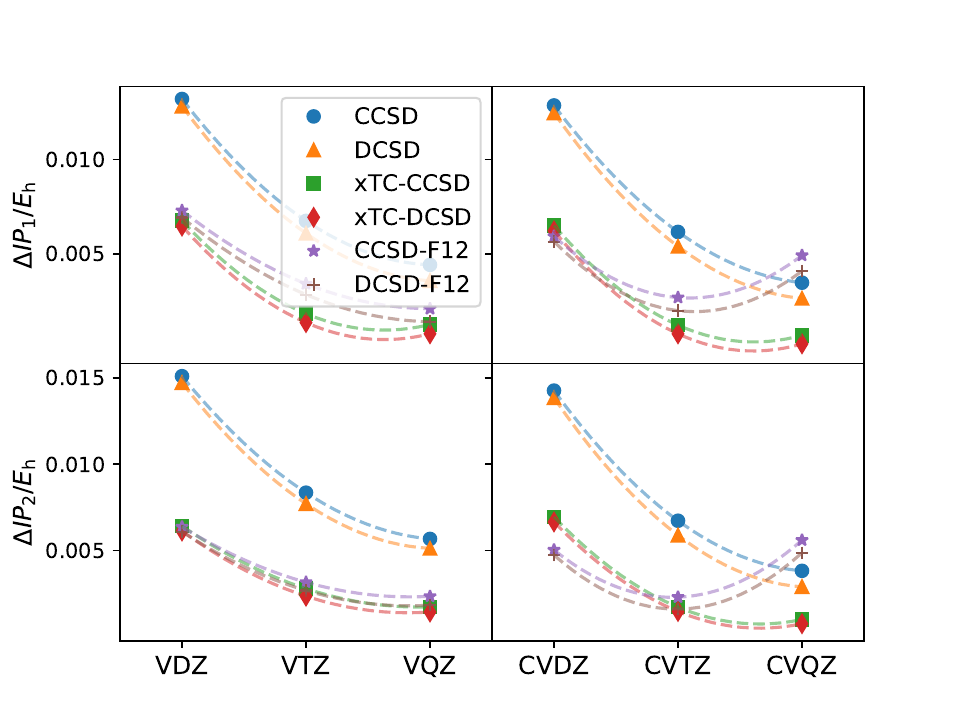}
    \caption{Basis set dependence of the mean absolute error in first ionisation potentials (top) and second ionisation potentials (bottom) computed with CCSD, DCSD, xTC-CCSD (B), xTC-DCSD(B), CCSD-F12a and DCSD-F12a. }%Dashed lines correspond to average errors in the ionisation potentials across all second row atoms. The dot-dashed lines in the bottom left panel exclude the anomalous Na$^+$ ionisation potential, recovering the trends observed in the other panels.}
    \label{fig:ip_bde}
\end{figure*}

\Cref{fig:ip_bde} shows the MAE of the ionisation potentials in the cc-pVXZ basis sets. Once again we observe that error decreases with basis set cardinality. However, due to error cancellation the performance of the core-valence basis sets is only marginally better.
We also note that as cardinality increases, so does the amount by which DCSD approaches outperform their CCSD counterparts. In this case xTC approaches converge faster than F12 methods with basis set size. The latter additionally display a large increase in MAE on going from cc-pCVTZ to cc-pCVQZ. In contrast, xTC-DCSD is near or within chemical accuracy for all systems at the quadruple zeta level, with or without using core-valence basis sets.

\subsection{Wavefunction compactification}
Another observed benefit of the transcorrelated approach is that it generally leads to more compact right eigenvectors of the Hamiltonian.\cite{Dobrautz2019,McArdle2020,Sokolov2020,Haupt2023, Dobrautz2024,Ammar2024} This corresponds to fewer determinants with large coefficients and may be easily ascertained for FCIQMC. In the past, a metric for compactification has been defined\cite{Haupt2023} as
\begin{equation}
    \xi = \frac{c_\text{HF}^\text{(TC)} - c_\text{HF}^\text{(non-TC)}}{1 - c_\text{HF}^\text{(non-TC)}},
\end{equation}
where $c_\text{HF}^\text{(TC)}$ and $c_\text{HF}^\text{(non-TC)}$ are the unsigned coefficients of the HF determinant in the normalised TC and non-TC wavefunctions, respectively.
\begin{table}[h!]
\caption{\small Compactification of FCIQMC atomic wavefunctions between non-TC and TC calculations, as measured by $\xi$. A value of 0 corresponds to no compactification, while 1 corresponds to a single-determinant wavefunction.}
\label{tab:compact}
\centering
\begin{tabular}{l|c|c|c|c|c|c}
&\multicolumn{3}{|c}{(A)}&\multicolumn{3}{|c}{(B)}\\
& cc-pVDZ & cc-pVTZ & cc-pVQZ& cc-pVDZ & cc-pVTZ & cc-pVQZ\\
\hline
Na& 0.374& 0.482& 0.435 & -0.240 &0.344&0.426\\
Mg& 0.001& 0.020& 0.070 & 0.084&0.106&0.157\\
Al& 0.007& 0.100& 0.062 & 0.131 &0.191&0.154\\
Si& 0.033& 0.072& 0.175& 0.177 &0.222&0.342\\
P & 0.079& 0.166& 0.265& 0.280 &0.273&0.559\\
S & 0.110& 0.161& 0.323& 0.368 & 0.334&0.576\\
Cl & 0.183& 0.267&0.408& 0.450& 0.416&0.671\\
Ar & 0.266& 0.429&0.487& 0.538 &0.509&0.827 \\
\hline
\end{tabular}
\end{table}
 $\xi > 0$ indicates a compactification of the wavefunction, with $\xi=1$ corresponding to a perfectly compact, single determinant wavefunction. A negative value of $\xi$ denotes a TC wavefunction that is less compact. We only observe this behaviour in the case of Na in the cc-pVDZ basis set. In this case, the coefficient of the HF determinant in the conventional FCIQMC calculation is $c_\text{HF} = 0.99975$, so the wavefunction has predominantly single-reference character. The potential for  compactification is very small in this case, but the structure of $\xi$ makes it sensitive to small changes in the HF coefficient in this scenario. The coefficients in the (A) and (B) cases are $c_\text{HF}^\text{A} = 0.99985$ and $c_\text{HF}^\text{B} = 0.99969$, which differ negligibly from $c_\text{HF}$, likely due to noise in the QMC procedure.

Values of $\xi$ for the second row atoms are shown in \cref{tab:compact}. In all cases, transcorrelation leads to some compactification, although the obtained $\xi$ are highly system-dependent and generally smaller than those reported for first row atoms.\cite{Haupt2023} The (B) set of results show larger degrees of compactification than the (A) set, which is unsurprising given the larger cutoffs allow the Jastrow factor to capture more correlation in this case. Most importantly, the increase in compactness of the eigenvector generally becomes more significant with increasing basis set size, which is promising for the applicability of methods such as FCIQMC.

\subsection{Frozen core approximation}
Even with the increased accuracy provided by the transcorrelated approach, it is clear that relatively large basis sets are needed to accurately describe the second row atoms, requiring increased computational resources. One approach to reduce these costs is to freeze the core electrons. The previous work of Christlmaier \textit{et al}\cite{Christlmaier2023} has shown that freezing the core electrons before computing the TC integrals  introduces significant errors in the final results. However, freezing the core orbitals after the TC calculation but before the post-Hartree--Fock correlation treatment only generates negligible errors.

\begin{table}
\caption{\small Mean absolute error (MAE), standard deviation (STD), root mean squared error (RMSD) and maximum error (ME) for ionisation energies computed with xTC-DCSD under different frozen core approximations in the cc-pVTZ basis. Na and Mg$^{+}$ ionisation potentials are excluded. All values are in mHartree.}
\label{tab:fc}
\centering

    \begin{tabular}
{l|c|c|c|c|c|c}

     \multirow{2}{*}{Method} & \multicolumn{3}{|c|}{$E_{\text{IP}_1}$}&\multicolumn{3}{|c}{$E_{\text{IP}_2}$} \\
     \cline{2-7}
         & MAE & RMSD & ME &MAE & RMSD & ME   \\
         \hline
     xTC-DCSD&0.9&1.2&3.1&2.0&1.8&4.2\\
     xTC-DCSD(He)&0.9&1.2&2.0&2.5&3.8&4.2\\
     xTC-DCSD(Ne)&0.9&1.2&2.5&2.7&4.2&3.6\\
     DCSD-F12a&3.0&3.6&5.6&2.8&3.4&5.4\\
     DCSD-F12a(He)&3.0&3.5&5.2&2.9&3.4&5.5\\
     DCSD-F12a(Ne)&3.5&3.9&5.6&4.5&4.8&6.3\\
     \hline
     \end{tabular}
\end{table}
For the second row atoms, there are two reasonable frozen core approximations, obtained by freezing either the He [$1s^2$] or the Ne [$1s^2 2s^2 2p^6$] core. The latter would lead to a more significant reduction in computation, but it cannot be applied to the ionisation of Na and Mg$^+$, as the $n=2$ shell becomes the valence shell for the resulting ions. We compare the quality of both approximations for ionisation potentials in \cref{tab:fc}, for the cc-pVTZ basis set. For the first IP, both approximations have negligible effect on the results, in both xTC and F12 approaches. For the second IP, freezing the Ne core introduces a slightly larger discrepancy from the all-electron calculations, although it remains in the sub-mHartree range for xTC. In contrast, the quality of the F12 results deteriorates significantly under this approximation. Frozen Ne core results are also shown for the cc-pVQZ basis set in \cref{tab:qz-ip-errors}, exhibiting the same trends, although in this case the F12 first ionisation potentials also deteriorate.

\section{Conclusion}
\label{sec:conclusion}

We investigated the use of transcorrelated methods for second row elements, using flexible, parametrised Jastrow factors and the highly effective xTC approximation. We found that in general xTC approaches lead to an approximately four-fold reduction in total energy errors relative to conventional approaches and a significant decrease in ionisation potential errors as well. Explicitly correlated F12a methods are more effective in reducing absolute errors, but xTC approaches become competitive for relative energies, in particular in the larger cc-pVQZ basis set and when using larger cutoff values in the Jastrow factor.

TC results show faster and more accurate convergence with basis set size than conventional approaches, as well as increased compactness of the wavefunction for most systems. Frozen core approximations for both the He and the Ne core can be used, introducing only minor discrepancies in the energies and, in general, we find that the cc-pVTZ basis set is sufficient to obtain chemically accurate ionisation potentials.

While the TC approach leads to significant accuracy improvements, calculations on the second row atoms still require relatively large basis sets. It will therefore be worthwhile to explore methods which may reduce the cost of calculations in these basis sets beyond the frozen core approximation, such as using effective core potentials.\cite{Simula2024}

\section*{Supplementary Material}
Individual energies for all species and methods considered can be found in the supplementary material.

\begin{acknowledgments}
We gratefully acknowledge financial support from the Max Planck Society.
\end{acknowledgments}

\section*{Data Availability Statement}
The data that support the findings of
this study are available within the
article and its supplementary material.

\nocite{*}
\bibliography{bibliography}% Produces the bibliography via BibTeX.

\begin{thebibliography}{74}%
\makeatletter
\providecommand \@ifxundefined [1]{%
 \@ifx{#1\undefined}
}%
\providecommand \@ifnum [1]{%
 \ifnum #1\expandafter \@firstoftwo
 \else \expandafter \@secondoftwo
 \fi
}%
\providecommand \@ifx [1]{%
 \ifx #1\expandafter \@firstoftwo
 \else \expandafter \@secondoftwo
 \fi
}%
\providecommand \natexlab [1]{#1}%
\providecommand \enquote  [1]{``#1''}%
\providecommand \bibnamefont  [1]{#1}%
\providecommand \bibfnamefont [1]{#1}%
\providecommand \citenamefont [1]{#1}%
\providecommand \href@noop [0]{\@secondoftwo}%
\providecommand \href [0]{\begingroup \@sanitize@url \@href}%
\providecommand \@href[1]{\@@startlink{#1}\@@href}%
\providecommand \@@href[1]{\endgroup#1\@@endlink}%
\providecommand \@sanitize@url [0]{\catcode `\\12\catcode `\$12\catcode `\&12\catcode `\#12\catcode `\^12\catcode `\_12\catcode `\%12\relax}%
\providecommand \@@startlink[1]{}%
\providecommand \@@endlink[0]{}%
\providecommand \url  [0]{\begingroup\@sanitize@url \@url }%
\providecommand \@url [1]{\endgroup\@href {#1}{\urlprefix }}%
\providecommand \urlprefix  [0]{URL }%
\providecommand \Eprint [0]{\href }%
\providecommand \doibase [0]{http://dx.doi.org/}%
\providecommand \selectlanguage [0]{\@gobble}%
\providecommand \bibinfo  [0]{\@secondoftwo}%
\providecommand \bibfield  [0]{\@secondoftwo}%
\providecommand \translation [1]{[#1]}%
\providecommand \BibitemOpen [0]{}%
\providecommand \bibitemStop [0]{}%
\providecommand \bibitemNoStop [0]{.\EOS\space}%
\providecommand \EOS [0]{\spacefactor3000\relax}%
\providecommand \BibitemShut  [1]{\csname bibitem#1\endcsname}%
\let\auto@bib@innerbib\@empty
%</preamble>
\bibitem [{\citenamefont {Kato}(1957)}]{Kato1957}%
  \BibitemOpen
  \bibfield  {author} {\bibinfo {author} {\bibfnamefont {T.}~\bibnamefont {Kato}},\ }\bibfield  {title} {\enquote {\bibinfo {title} {On the eigenfunctions of many-particle systems in quantum mechanics},}\ }\href {\doibase https://doi.org/10.1002/cpa.3160100201} {\bibfield  {journal} {\bibinfo  {journal} {Commun. Pure Appl. Math.}\ }\textbf {\bibinfo {volume} {10}},\ \bibinfo {pages} {151--177} (\bibinfo {year} {1957})}\BibitemShut {NoStop}%
\bibitem [{\citenamefont {Hylleraas}(1928)}]{Hylleraas1928}%
  \BibitemOpen
  \bibfield  {author} {\bibinfo {author} {\bibfnamefont {E.~A.}\ \bibnamefont {Hylleraas}},\ }\bibfield  {title} {\enquote {\bibinfo {title} {Über den grundzustand des heliumatoms},}\ }\href {\doibase https://doi.org/10.1007/BF01340013} {\bibfield  {journal} {\bibinfo  {journal} {Z. Physik}\ }\textbf {\bibinfo {volume} {48}},\ \bibinfo {pages} {469--494} (\bibinfo {year} {1928})}\BibitemShut {NoStop}%
\bibitem [{\citenamefont {Hylleraas}(1929)}]{Hylleraas1929}%
  \BibitemOpen
  \bibfield  {author} {\bibinfo {author} {\bibfnamefont {E.~A.}\ \bibnamefont {Hylleraas}},\ }\bibfield  {title} {\enquote {\bibinfo {title} {Neue berechnung der energie des heliums im grundzustande, sowie des tiefsten terms von ortho-helium},}\ }\href {\doibase https://doi.org/10.1007/BF01375457} {\bibfield  {journal} {\bibinfo  {journal} {Z. Physik}\ }\textbf {\bibinfo {volume} {54}},\ \bibinfo {pages} {347--366} (\bibinfo {year} {1929})}\BibitemShut {NoStop}%
\bibitem [{\citenamefont {Hylleraas}(1920)}]{Hylleraas1930}%
  \BibitemOpen
  \bibfield  {author} {\bibinfo {author} {\bibfnamefont {E.~A.}\ \bibnamefont {Hylleraas}},\ }\bibfield  {title} {\enquote {\bibinfo {title} {\"uber den grundterm der zweielektronenprobleme von h-, he, li+, be++ usw.}}\ }\href {\doibase https://doi.org/10.1007/BF01397032} {\bibfield  {journal} {\bibinfo  {journal} {Z. Physik}\ }\textbf {\bibinfo {volume} {65}},\ \bibinfo {pages} {209--225} (\bibinfo {year} {1920})}\BibitemShut {NoStop}%
\bibitem [{\citenamefont {Kutzelnigg}(1985)}]{Kutzelnigg1985}%
  \BibitemOpen
  \bibfield  {author} {\bibinfo {author} {\bibfnamefont {W.}~\bibnamefont {Kutzelnigg}},\ }\bibfield  {title} {\enquote {\bibinfo {title} {r12-dependent terms in the wave function as closed sums of partial wave amplitudes for large l},}\ }\href {\doibase 10.1007/BF00527669/METRICS} {\bibfield  {journal} {\bibinfo  {journal} {Theoret. Chim. Acta}\ }\textbf {\bibinfo {volume} {68}},\ \bibinfo {pages} {445--469} (\bibinfo {year} {1985})}\BibitemShut {NoStop}%
\bibitem [{\citenamefont {Klopper}\ and\ \citenamefont {Kutzelnigg}(1987)}]{Klopper1987}%
  \BibitemOpen
  \bibfield  {author} {\bibinfo {author} {\bibfnamefont {W.}~\bibnamefont {Klopper}}\ and\ \bibinfo {author} {\bibfnamefont {W.}~\bibnamefont {Kutzelnigg}},\ }\bibfield  {title} {\enquote {\bibinfo {title} {Møller-plesset calculations taking care of the correlation cusp},}\ }\href {\doibase 10.1016/0009-2614(87)80005-2} {\bibfield  {journal} {\bibinfo  {journal} {Chem. Phys. Lett.}\ }\textbf {\bibinfo {volume} {134}},\ \bibinfo {pages} {17--22} (\bibinfo {year} {1987})}\BibitemShut {NoStop}%
\bibitem [{\citenamefont {Klopper}\ and\ \citenamefont {Kutzelnigg}(1990)}]{Klopper1990}%
  \BibitemOpen
  \bibfield  {author} {\bibinfo {author} {\bibfnamefont {W.}~\bibnamefont {Klopper}}\ and\ \bibinfo {author} {\bibfnamefont {W.}~\bibnamefont {Kutzelnigg}},\ }\bibfield  {title} {\enquote {\bibinfo {title} {Mp2-r12 calculations on the relative stability of carbocations},}\ }\href {\doibase 10.1021/J100377A040/ASSET/J100377A040.FP.PNG_V03} {\bibfield  {journal} {\bibinfo  {journal} {J. Phys. Chem.}\ }\textbf {\bibinfo {volume} {94}},\ \bibinfo {pages} {5625--5630} (\bibinfo {year} {1990})}\BibitemShut {NoStop}%
\bibitem [{\citenamefont {Kutzelnigg}\ and\ \citenamefont {Klopper}(1991)}]{Kutzelnigg1991}%
  \BibitemOpen
  \bibfield  {author} {\bibinfo {author} {\bibfnamefont {W.}~\bibnamefont {Kutzelnigg}}\ and\ \bibinfo {author} {\bibfnamefont {W.}~\bibnamefont {Klopper}},\ }\bibfield  {title} {\enquote {\bibinfo {title} {Wave functions with terms linear in the interelectronic coordinates to take care of the correlation cusp. i. general theory},}\ }\href {\doibase 10.1063/1.459921} {\bibfield  {journal} {\bibinfo  {journal} {J. Chem. Phys.}\ }\textbf {\bibinfo {volume} {94}},\ \bibinfo {pages} {1985--2001} (\bibinfo {year} {1991})}\BibitemShut {NoStop}%
\bibitem [{\citenamefont {Noga}\ and\ \citenamefont {Kutzelnigg}(1994)}]{Noga1994}%
  \BibitemOpen
  \bibfield  {author} {\bibinfo {author} {\bibfnamefont {J.}~\bibnamefont {Noga}}\ and\ \bibinfo {author} {\bibfnamefont {W.}~\bibnamefont {Kutzelnigg}},\ }\bibfield  {title} {\enquote {\bibinfo {title} {Coupled cluster theory that takes care of the correlation cusp by inclusion of linear terms in the interelectronic coordinates},}\ }\href {\doibase 10.1063/1.468266} {\bibfield  {journal} {\bibinfo  {journal} {J. Chem. Phys.}\ }\textbf {\bibinfo {volume} {101}},\ \bibinfo {pages} {7738--7762} (\bibinfo {year} {1994})}\BibitemShut {NoStop}%
\bibitem [{\citenamefont {Noga}, \citenamefont {Klopper},\ and\ \citenamefont {Kutzelnigg}(1997)}]{Noga1997}%
  \BibitemOpen
  \bibfield  {author} {\bibinfo {author} {\bibfnamefont {J.}~\bibnamefont {Noga}}, \bibinfo {author} {\bibfnamefont {W.}~\bibnamefont {Klopper}}, \ and\ \bibinfo {author} {\bibfnamefont {W.}~\bibnamefont {Kutzelnigg}},\ }\bibfield  {title} {\enquote {\bibinfo {title} {Cc-r12: An explicitly correlated coupled-cluster theory},}\ }\href {\doibase 10.1142/9789812819529_0001} {\bibfield  {journal} {\bibinfo  {journal} {Recent Advances in Coupled-Cluster Methods}\ ,\ \bibinfo {pages} {1--48}} (\bibinfo {year} {1997})}\BibitemShut {NoStop}%
\bibitem [{\citenamefont {Ten-no}(2004)}]{Ten-no2004}%
  \BibitemOpen
  \bibfield  {author} {\bibinfo {author} {\bibfnamefont {S.}~\bibnamefont {Ten-no}},\ }\bibfield  {title} {\enquote {\bibinfo {title} {Initiation of explicitly correlated slater-type geminal theory},}\ }\href {\doibase 10.1016/J.CPLETT.2004.09.041} {\bibfield  {journal} {\bibinfo  {journal} {Chem. Phys. Lett.}\ }\textbf {\bibinfo {volume} {398}},\ \bibinfo {pages} {56--61} (\bibinfo {year} {2004})}\BibitemShut {NoStop}%
\bibitem [{\citenamefont {Valeev}(2004)}]{Valeev2004}%
  \BibitemOpen
  \bibfield  {author} {\bibinfo {author} {\bibfnamefont {E.~F.}\ \bibnamefont {Valeev}},\ }\bibfield  {title} {\enquote {\bibinfo {title} {Improving on the resolution of the identity in linear r12 ab initio theories},}\ }\href {\doibase 10.1016/J.CPLETT.2004.07.061} {\bibfield  {journal} {\bibinfo  {journal} {Chem. Phys. Lett.}\ }\textbf {\bibinfo {volume} {395}},\ \bibinfo {pages} {190--195} (\bibinfo {year} {2004})}\BibitemShut {NoStop}%
\bibitem [{\citenamefont {Kedžuch}, \citenamefont {Milko},\ and\ \citenamefont {Noga}(2005)}]{Kedzuch2005}%
  \BibitemOpen
  \bibfield  {author} {\bibinfo {author} {\bibfnamefont {S.}~\bibnamefont {Kedžuch}}, \bibinfo {author} {\bibfnamefont {M.}~\bibnamefont {Milko}}, \ and\ \bibinfo {author} {\bibfnamefont {J.}~\bibnamefont {Noga}},\ }\bibfield  {title} {\enquote {\bibinfo {title} {Alternative formulation of the matrix elements in mp2-r12 theory},}\ }\href {\doibase 10.1002/QUA.20744} {\bibfield  {journal} {\bibinfo  {journal} {Int. J. Quantum Chem.}\ }\textbf {\bibinfo {volume} {105}},\ \bibinfo {pages} {929--936} (\bibinfo {year} {2005})}\BibitemShut {NoStop}%
\bibitem [{\citenamefont {Adler}, \citenamefont {Knizia},\ and\ \citenamefont {Werner}(2007)}]{Adler2007}%
  \BibitemOpen
  \bibfield  {author} {\bibinfo {author} {\bibfnamefont {T.~B.}\ \bibnamefont {Adler}}, \bibinfo {author} {\bibfnamefont {G.}~\bibnamefont {Knizia}}, \ and\ \bibinfo {author} {\bibfnamefont {H.~J.}\ \bibnamefont {Werner}},\ }\bibfield  {title} {\enquote {\bibinfo {title} {A simple and efficient ccsd(t)-f12 approximation},}\ }\href {\doibase 10.1063/1.2817618/918519} {\bibfield  {journal} {\bibinfo  {journal} {J. Chem. Phys.}\ }\textbf {\bibinfo {volume} {127}},\ \bibinfo {pages} {221106} (\bibinfo {year} {2007})}\BibitemShut {NoStop}%
\bibitem [{\citenamefont {Knizia}\ and\ \citenamefont {Werner}(2008)}]{Knizia2008}%
  \BibitemOpen
  \bibfield  {author} {\bibinfo {author} {\bibfnamefont {G.}~\bibnamefont {Knizia}}\ and\ \bibinfo {author} {\bibfnamefont {H.~J.}\ \bibnamefont {Werner}},\ }\bibfield  {title} {\enquote {\bibinfo {title} {Explicitly correlated rmp2 for high-spin open-shell reference states},}\ }\href {\doibase 10.1063/1.2889388/71081} {\bibfield  {journal} {\bibinfo  {journal} {J. Chem. Phys.}\ }\textbf {\bibinfo {volume} {128}},\ \bibinfo {pages} {154103} (\bibinfo {year} {2008})}\BibitemShut {NoStop}%
\bibitem [{\citenamefont {Knizia}, \citenamefont {Adler},\ and\ \citenamefont {Werner}(2009)}]{Knizia2009}%
  \BibitemOpen
  \bibfield  {author} {\bibinfo {author} {\bibfnamefont {G.}~\bibnamefont {Knizia}}, \bibinfo {author} {\bibfnamefont {T.~B.}\ \bibnamefont {Adler}}, \ and\ \bibinfo {author} {\bibfnamefont {H.~J.}\ \bibnamefont {Werner}},\ }\bibfield  {title} {\enquote {\bibinfo {title} {Simplified ccsd(t)-f12 methods: Theory and benchmarks},}\ }\href {\doibase 10.1063/1.3054300/908511} {\bibfield  {journal} {\bibinfo  {journal} {J. Chem. Phys.}\ }\textbf {\bibinfo {volume} {130}},\ \bibinfo {pages} {54104} (\bibinfo {year} {2009})}\BibitemShut {NoStop}%
\bibitem [{\citenamefont {Hättig}, \citenamefont {Tew},\ and\ \citenamefont {Köhn}(2010)}]{Hattig2010}%
  \BibitemOpen
  \bibfield  {author} {\bibinfo {author} {\bibfnamefont {C.}~\bibnamefont {Hättig}}, \bibinfo {author} {\bibfnamefont {D.~P.}\ \bibnamefont {Tew}}, \ and\ \bibinfo {author} {\bibfnamefont {A.}~\bibnamefont {Köhn}},\ }\bibfield  {title} {\enquote {\bibinfo {title} {Communications: Accurate and efficient approximations to explicitly correlated coupled-cluster singles and doubles, ccsd-f12},}\ }\href {\doibase 10.1063/1.3442368/71472} {\bibfield  {journal} {\bibinfo  {journal} {J. Chem. Phys.}\ }\textbf {\bibinfo {volume} {132}},\ \bibinfo {pages} {231102} (\bibinfo {year} {2010})}\BibitemShut {NoStop}%
\bibitem [{\citenamefont {Kong}, \citenamefont {Bischoff},\ and\ \citenamefont {Valeev}(2012)}]{Kong2012}%
  \BibitemOpen
  \bibfield  {author} {\bibinfo {author} {\bibfnamefont {L.}~\bibnamefont {Kong}}, \bibinfo {author} {\bibfnamefont {F.~A.}\ \bibnamefont {Bischoff}}, \ and\ \bibinfo {author} {\bibfnamefont {E.~F.}\ \bibnamefont {Valeev}},\ }\bibfield  {title} {\enquote {\bibinfo {title} {Explicitly correlated r12/f12 methods for electronic structure},}\ }\href {\doibase 10.1021/cr200204r} {\bibfield  {journal} {\bibinfo  {journal} {Chem. Rev.}\ }\textbf {\bibinfo {volume} {112}},\ \bibinfo {pages} {75--107} (\bibinfo {year} {2012})}\BibitemShut {NoStop}%
\bibitem [{\citenamefont {Boys}\ and\ \citenamefont {Handy}(1969)}]{Boys1969}%
  \BibitemOpen
  \bibfield  {author} {\bibinfo {author} {\bibfnamefont {S.~F.}\ \bibnamefont {Boys}}\ and\ \bibinfo {author} {\bibfnamefont {N.~C.}\ \bibnamefont {Handy}},\ }\bibfield  {title} {\enquote {\bibinfo {title} {The determination of energies and wavefunctions with full electronic correlation},}\ }\href {\doibase http://doi.org/10.1098/rspa.1969.0061} {\bibfield  {journal} {\bibinfo  {journal} {Proc. R. Soc. Lond. A}\ }\textbf {\bibinfo {volume} {310}},\ \bibinfo {pages} {43--61} (\bibinfo {year} {1969})}\BibitemShut {NoStop}%
\bibitem [{\citenamefont {Handy}(1969)}]{Handy1969}%
  \BibitemOpen
  \bibfield  {author} {\bibinfo {author} {\bibfnamefont {N.~C.}\ \bibnamefont {Handy}},\ }\bibfield  {title} {\enquote {\bibinfo {title} {{Energies and Expectation Values for Be by the Transcorrelated Method}},}\ }\href {\doibase 10.1063/1.1672496} {\bibfield  {journal} {\bibinfo  {journal} {J. Chem. Phys.}\ }\textbf {\bibinfo {volume} {51}},\ \bibinfo {pages} {3205--3212} (\bibinfo {year} {1969})}\BibitemShut {NoStop}%
\bibitem [{\citenamefont {Handy}(1971)}]{Handy1971}%
  \BibitemOpen
  \bibfield  {author} {\bibinfo {author} {\bibfnamefont {N.}~\bibnamefont {Handy}},\ }\bibfield  {title} {\enquote {\bibinfo {title} {On the minimization of the variance of the transcorrelated hamiltonian},}\ }\href {\doibase 10.1080/00268977100101961} {\bibfield  {journal} {\bibinfo  {journal} {Mol. Phys.}\ }\textbf {\bibinfo {volume} {21}},\ \bibinfo {pages} {817--828} (\bibinfo {year} {1971})}\BibitemShut {NoStop}%
\bibitem [{\citenamefont {Jastrow}(1955)}]{Jastrow1955}%
  \BibitemOpen
  \bibfield  {author} {\bibinfo {author} {\bibfnamefont {R.}~\bibnamefont {Jastrow}},\ }\bibfield  {title} {\enquote {\bibinfo {title} {Many-body problem with strong forces},}\ }\href {\doibase 10.1103/PhysRev.98.1479} {\bibfield  {journal} {\bibinfo  {journal} {Phys. Rev.}\ }\textbf {\bibinfo {volume} {98}},\ \bibinfo {pages} {1479--1484} (\bibinfo {year} {1955})}\BibitemShut {NoStop}%
\bibitem [{\citenamefont {Nooijen}\ and\ \citenamefont {Bartlett}(1998)}]{Nooijen1998}%
  \BibitemOpen
  \bibfield  {author} {\bibinfo {author} {\bibfnamefont {M.}~\bibnamefont {Nooijen}}\ and\ \bibinfo {author} {\bibfnamefont {R.~J.}\ \bibnamefont {Bartlett}},\ }\bibfield  {title} {\enquote {\bibinfo {title} {{Elimination of Coulombic infinities through transformation of the Hamiltonian}},}\ }\href {\doibase 10.1063/1.477485} {\bibfield  {journal} {\bibinfo  {journal} {J. Chem. Phys.}\ }\textbf {\bibinfo {volume} {109}},\ \bibinfo {pages} {8232--8240} (\bibinfo {year} {1998})}\BibitemShut {NoStop}%
\bibitem [{\citenamefont {Ten-no}(2000{\natexlab{a}})}]{Tenno2000}%
  \BibitemOpen
  \bibfield  {author} {\bibinfo {author} {\bibfnamefont {S.}~\bibnamefont {Ten-no}},\ }\bibfield  {title} {\enquote {\bibinfo {title} {Three-electron integral evaluation in the transcorrelated method using a frozen gaussian geminal},}\ }\href {\doibase https://doi.org/10.1016/S0009-2614(00)01067-8} {\bibfield  {journal} {\bibinfo  {journal} {Chem. Phys. Lett.}\ }\textbf {\bibinfo {volume} {330}},\ \bibinfo {pages} {175--179} (\bibinfo {year} {2000}{\natexlab{a}})}\BibitemShut {NoStop}%
\bibitem [{\citenamefont {Ten-no}(2000{\natexlab{b}})}]{Tenno2000b}%
  \BibitemOpen
  \bibfield  {author} {\bibinfo {author} {\bibfnamefont {S.}~\bibnamefont {Ten-no}},\ }\bibfield  {title} {\enquote {\bibinfo {title} {A feasible transcorrelated method for treating electronic cusps using a frozen gaussian geminal},}\ }\href {\doibase https://doi.org/10.1016/S0009-2614(00)01066-6} {\bibfield  {journal} {\bibinfo  {journal} {Chem. Phys. Lett.}\ }\textbf {\bibinfo {volume} {330}},\ \bibinfo {pages} {169--174} (\bibinfo {year} {2000}{\natexlab{b}})}\BibitemShut {NoStop}%
\bibitem [{\citenamefont {Hino}, \citenamefont {Tanimura},\ and\ \citenamefont {Ten-no}(2002)}]{Hino2002}%
  \BibitemOpen
  \bibfield  {author} {\bibinfo {author} {\bibfnamefont {O.}~\bibnamefont {Hino}}, \bibinfo {author} {\bibfnamefont {Y.}~\bibnamefont {Tanimura}}, \ and\ \bibinfo {author} {\bibfnamefont {S.}~\bibnamefont {Ten-no}},\ }\bibfield  {title} {\enquote {\bibinfo {title} {Application of the transcorrelated hamiltonian to the linearized coupled cluster singles and doubles model},}\ }\href {\doibase https://doi.org/10.1016/S0009-2614(02)00042-8} {\bibfield  {journal} {\bibinfo  {journal} {Chem. Phys. Lett.}\ }\textbf {\bibinfo {volume} {353}},\ \bibinfo {pages} {317--323} (\bibinfo {year} {2002})}\BibitemShut {NoStop}%
\bibitem [{\citenamefont {Umezawa}\ and\ \citenamefont {Tsuneyuki}(2003)}]{Umezawa2003}%
  \BibitemOpen
  \bibfield  {author} {\bibinfo {author} {\bibfnamefont {N.}~\bibnamefont {Umezawa}}\ and\ \bibinfo {author} {\bibfnamefont {S.}~\bibnamefont {Tsuneyuki}},\ }\bibfield  {title} {\enquote {\bibinfo {title} {{Transcorrelated method for electronic systems coupled with variational Monte Carlo calculation}},}\ }\href {\doibase 10.1063/1.1617274} {\bibfield  {journal} {\bibinfo  {journal} {J. Chem. Phys.}\ }\textbf {\bibinfo {volume} {119}},\ \bibinfo {pages} {10015--10031} (\bibinfo {year} {2003})}\BibitemShut {NoStop}%
\bibitem [{\citenamefont {Umezawa}\ and\ \citenamefont {Tsuneyuki}(2004)}]{Umezawa2004}%
  \BibitemOpen
  \bibfield  {author} {\bibinfo {author} {\bibfnamefont {N.}~\bibnamefont {Umezawa}}\ and\ \bibinfo {author} {\bibfnamefont {S.}~\bibnamefont {Tsuneyuki}},\ }\bibfield  {title} {\enquote {\bibinfo {title} {Ground-state correlation energy for the homogeneous electron gas calculated by the transcorrelated method},}\ }\href {\doibase 10.1103/PhysRevB.69.165102} {\bibfield  {journal} {\bibinfo  {journal} {Phys. Rev. B}\ }\textbf {\bibinfo {volume} {69}},\ \bibinfo {pages} {165102} (\bibinfo {year} {2004})}\BibitemShut {NoStop}%
\bibitem [{\citenamefont {Sakuma}\ and\ \citenamefont {Tsuneyuki}(2006)}]{Sakuma2006}%
  \BibitemOpen
  \bibfield  {author} {\bibinfo {author} {\bibfnamefont {R.}~\bibnamefont {Sakuma}}\ and\ \bibinfo {author} {\bibfnamefont {S.}~\bibnamefont {Tsuneyuki}},\ }\bibfield  {title} {\enquote {\bibinfo {title} {Electronic structure calculations of solids with a similarity-transformed hamiltonian},}\ }\href {\doibase 10.1143/JPSJ.75.103705} {\bibfield  {journal} {\bibinfo  {journal} {J. Phys. Soc. Jpn.}\ }\textbf {\bibinfo {volume} {75}},\ \bibinfo {pages} {103705} (\bibinfo {year} {2006})}\BibitemShut {NoStop}%
\bibitem [{\citenamefont {Tsuneyuki}(2008)}]{Tsuneyuki2008}%
  \BibitemOpen
  \bibfield  {author} {\bibinfo {author} {\bibfnamefont {S.}~\bibnamefont {Tsuneyuki}},\ }\bibfield  {title} {\enquote {\bibinfo {title} {{Transcorrelated Method: Another Possible Way towards Electronic Structure Calculation of Solids}},}\ }\href {\doibase 10.1143/PTPS.176.134} {\bibfield  {journal} {\bibinfo  {journal} {Prog. Theor. Phys. Supp.}\ }\textbf {\bibinfo {volume} {176}},\ \bibinfo {pages} {134--142} (\bibinfo {year} {2008})}\BibitemShut {NoStop}%
\bibitem [{\citenamefont {Luo}\ and\ \citenamefont {Alavi}(2018)}]{Luo2018}%
  \BibitemOpen
  \bibfield  {author} {\bibinfo {author} {\bibfnamefont {H.}~\bibnamefont {Luo}}\ and\ \bibinfo {author} {\bibfnamefont {A.}~\bibnamefont {Alavi}},\ }\bibfield  {title} {\enquote {\bibinfo {title} {Combining the transcorrelated method with full configuration interaction quantum monte carlo: Application to the homogeneous electron gas},}\ }\href {\doibase 10.1021/acs.jctc.7b01257} {\bibfield  {journal} {\bibinfo  {journal} {J. Chem. Theory Comput.}\ }\textbf {\bibinfo {volume} {14}},\ \bibinfo {pages} {1403--1411} (\bibinfo {year} {2018})}\BibitemShut {NoStop}%
\bibitem [{\citenamefont {Cohen}\ \emph {et~al.}(2019)\citenamefont {Cohen}, \citenamefont {Luo}, \citenamefont {Guther}, \citenamefont {Dobrautz}, \citenamefont {Tew},\ and\ \citenamefont {Alavi}}]{Cohen2019}%
  \BibitemOpen
  \bibfield  {author} {\bibinfo {author} {\bibfnamefont {A.~J.}\ \bibnamefont {Cohen}}, \bibinfo {author} {\bibfnamefont {H.}~\bibnamefont {Luo}}, \bibinfo {author} {\bibfnamefont {K.}~\bibnamefont {Guther}}, \bibinfo {author} {\bibfnamefont {W.}~\bibnamefont {Dobrautz}}, \bibinfo {author} {\bibfnamefont {D.~P.}\ \bibnamefont {Tew}}, \ and\ \bibinfo {author} {\bibfnamefont {A.}~\bibnamefont {Alavi}},\ }\bibfield  {title} {\enquote {\bibinfo {title} {{Similarity transformation of the electronic Schrödinger equation via Jastrow factorization}},}\ }\href {\doibase 10.1063/1.5116024} {\bibfield  {journal} {\bibinfo  {journal} {J. Chem. Phys.}\ }\textbf {\bibinfo {volume} {151}},\ \bibinfo {pages} {061101} (\bibinfo {year} {2019})}\BibitemShut {NoStop}%
\bibitem [{\citenamefont {Dobrautz}\ \emph {et~al.}(2022)\citenamefont {Dobrautz}, \citenamefont {Cohen}, \citenamefont {Alavi},\ and\ \citenamefont {Giner}}]{Dobrautz2022}%
  \BibitemOpen
  \bibfield  {author} {\bibinfo {author} {\bibfnamefont {W.}~\bibnamefont {Dobrautz}}, \bibinfo {author} {\bibfnamefont {A.~J.}\ \bibnamefont {Cohen}}, \bibinfo {author} {\bibfnamefont {A.}~\bibnamefont {Alavi}}, \ and\ \bibinfo {author} {\bibfnamefont {E.}~\bibnamefont {Giner}},\ }\bibfield  {title} {\enquote {\bibinfo {title} {Performance of a one-parameter correlation factor for transcorrelation: Study on a series of second row atomic and molecular systems},}\ }\href {\doibase 10.1063/5.0088981/2841332} {\bibfield  {journal} {\bibinfo  {journal} {J. Chem. Phys.}\ }\textbf {\bibinfo {volume} {156}},\ \bibinfo {pages} {234108} (\bibinfo {year} {2022})}\BibitemShut {NoStop}%
\bibitem [{\citenamefont {Ammar}, \citenamefont {Scemama},\ and\ \citenamefont {Giner}(2022)}]{Ammar2022}%
  \BibitemOpen
  \bibfield  {author} {\bibinfo {author} {\bibfnamefont {A.}~\bibnamefont {Ammar}}, \bibinfo {author} {\bibfnamefont {A.}~\bibnamefont {Scemama}}, \ and\ \bibinfo {author} {\bibfnamefont {E.}~\bibnamefont {Giner}},\ }\bibfield  {title} {\enquote {\bibinfo {title} {Extension of selected configuration interaction for transcorrelated methods},}\ }\href {\doibase 10.1063/5.0115524/2841826} {\bibfield  {journal} {\bibinfo  {journal} {J. Chem. Phys.}\ }\textbf {\bibinfo {volume} {157}},\ \bibinfo {pages} {134107} (\bibinfo {year} {2022})}\BibitemShut {NoStop}%
\bibitem [{\citenamefont {Ammar}, \citenamefont {Scemama},\ and\ \citenamefont {Giner}(2023{\natexlab{a}})}]{Ammar2023}%
  \BibitemOpen
  \bibfield  {author} {\bibinfo {author} {\bibfnamefont {A.}~\bibnamefont {Ammar}}, \bibinfo {author} {\bibfnamefont {A.}~\bibnamefont {Scemama}}, \ and\ \bibinfo {author} {\bibfnamefont {E.}~\bibnamefont {Giner}},\ }\bibfield  {title} {\enquote {\bibinfo {title} {Biorthonormal orbital optimization with a cheap core-electron-free three-body correlation factor for quantum monte carlo and transcorrelation},}\ }\href {\doibase 10.1021/ACS.JCTC.3C00257/ASSET/IMAGES/LARGE/CT3C00257_0007.JPEG} {\bibfield  {journal} {\bibinfo  {journal} {J. Chem. Theory Comput.}\ }\textbf {\bibinfo {volume} {19}},\ \bibinfo {pages} {4883--4896} (\bibinfo {year} {2023}{\natexlab{a}})}\BibitemShut {NoStop}%
\bibitem [{\citenamefont {Ammar}, \citenamefont {Scemama},\ and\ \citenamefont {Giner}(2023{\natexlab{b}})}]{Ammar2023b}%
  \BibitemOpen
  \bibfield  {author} {\bibinfo {author} {\bibfnamefont {A.}~\bibnamefont {Ammar}}, \bibinfo {author} {\bibfnamefont {A.}~\bibnamefont {Scemama}}, \ and\ \bibinfo {author} {\bibfnamefont {E.}~\bibnamefont {Giner}},\ }\bibfield  {title} {\enquote {\bibinfo {title} {Transcorrelated selected configuration interaction in a bi-orthonormal basis and with a cheap three-body correlation factor},}\ }\href {\doibase 10.1063/5.0163831/2912013} {\bibfield  {journal} {\bibinfo  {journal} {J. Chem. Phys.}\ }\textbf {\bibinfo {volume} {159}},\ \bibinfo {pages} {114121} (\bibinfo {year} {2023}{\natexlab{b}})}\BibitemShut {NoStop}%
\bibitem [{\citenamefont {Lee}\ and\ \citenamefont {Thom}(2023)}]{Lee2023}%
  \BibitemOpen
  \bibfield  {author} {\bibinfo {author} {\bibfnamefont {N.}~\bibnamefont {Lee}}\ and\ \bibinfo {author} {\bibfnamefont {A.~J.~W.}\ \bibnamefont {Thom}},\ }\bibfield  {title} {\enquote {\bibinfo {title} {Studies on the transcorrelated method},}\ }\href {\doibase 10.1021/acs.jctc.3c00046} {\bibfield  {journal} {\bibinfo  {journal} {J. Chem. Theory Comput.}\ }\textbf {\bibinfo {volume} {19}},\ \bibinfo {pages} {5743--5759} (\bibinfo {year} {2023})}\BibitemShut {NoStop}%
\bibitem [{\citenamefont {Ammar}\ \emph {et~al.}(2024)\citenamefont {Ammar}, \citenamefont {Scemama}, \citenamefont {Loos},\ and\ \citenamefont {Giner}}]{Ammar2024}%
  \BibitemOpen
  \bibfield  {author} {\bibinfo {author} {\bibfnamefont {A.}~\bibnamefont {Ammar}}, \bibinfo {author} {\bibfnamefont {A.}~\bibnamefont {Scemama}}, \bibinfo {author} {\bibfnamefont {P.-F.}\ \bibnamefont {Loos}}, \ and\ \bibinfo {author} {\bibfnamefont {E.}~\bibnamefont {Giner}},\ }\bibfield  {title} {\enquote {\bibinfo {title} {{Compactification of determinant expansions via transcorrelation}},}\ }\href {\doibase 10.1063/5.0217650} {\bibfield  {journal} {\bibinfo  {journal} {J. Chem. Phys.}\ }\textbf {\bibinfo {volume} {161}},\ \bibinfo {pages} {084104} (\bibinfo {year} {2024})}\BibitemShut {NoStop}%
\bibitem [{\citenamefont {Dobrautz}, \citenamefont {Luo},\ and\ \citenamefont {Alavi}(2019)}]{Dobrautz2019}%
  \BibitemOpen
  \bibfield  {author} {\bibinfo {author} {\bibfnamefont {W.}~\bibnamefont {Dobrautz}}, \bibinfo {author} {\bibfnamefont {H.}~\bibnamefont {Luo}}, \ and\ \bibinfo {author} {\bibfnamefont {A.}~\bibnamefont {Alavi}},\ }\bibfield  {title} {\enquote {\bibinfo {title} {Compact numerical solutions to the two-dimensional repulsive hubbard model obtained via nonunitary similarity transformations},}\ }\href {\doibase 10.1103/PhysRevB.99.075119} {\bibfield  {journal} {\bibinfo  {journal} {Phys. Rev. B}\ }\textbf {\bibinfo {volume} {99}},\ \bibinfo {pages} {075119} (\bibinfo {year} {2019})}\BibitemShut {NoStop}%
\bibitem [{\citenamefont {Liao}\ \emph {et~al.}(2021)\citenamefont {Liao}, \citenamefont {Schraivogel}, \citenamefont {Luo}, \citenamefont {Kats},\ and\ \citenamefont {Alavi}}]{Liao2021}%
  \BibitemOpen
  \bibfield  {author} {\bibinfo {author} {\bibfnamefont {K.}~\bibnamefont {Liao}}, \bibinfo {author} {\bibfnamefont {T.}~\bibnamefont {Schraivogel}}, \bibinfo {author} {\bibfnamefont {H.}~\bibnamefont {Luo}}, \bibinfo {author} {\bibfnamefont {D.}~\bibnamefont {Kats}}, \ and\ \bibinfo {author} {\bibfnamefont {A.}~\bibnamefont {Alavi}},\ }\bibfield  {title} {\enquote {\bibinfo {title} {Towards efficient and accurate ab initio solutions to periodic systems via transcorrelation and coupled cluster theory},}\ }\href {\doibase 10.1103/PhysRevResearch.3.033072} {\bibfield  {journal} {\bibinfo  {journal} {Phys. Rev. Res.}\ }\textbf {\bibinfo {volume} {3}},\ \bibinfo {pages} {033072} (\bibinfo {year} {2021})}\BibitemShut {NoStop}%
\bibitem [{\citenamefont {Schraivogel}\ \emph {et~al.}(2021)\citenamefont {Schraivogel}, \citenamefont {Cohen}, \citenamefont {Alavi},\ and\ \citenamefont {Kats}}]{Schraivogel2021}%
  \BibitemOpen
  \bibfield  {author} {\bibinfo {author} {\bibfnamefont {T.}~\bibnamefont {Schraivogel}}, \bibinfo {author} {\bibfnamefont {A.~J.}\ \bibnamefont {Cohen}}, \bibinfo {author} {\bibfnamefont {A.}~\bibnamefont {Alavi}}, \ and\ \bibinfo {author} {\bibfnamefont {D.}~\bibnamefont {Kats}},\ }\bibfield  {title} {\enquote {\bibinfo {title} {{Transcorrelated coupled cluster methods}},}\ }\href {\doibase 10.1063/5.0072495} {\bibfield  {journal} {\bibinfo  {journal} {J. Chem. Phys.}\ }\textbf {\bibinfo {volume} {155}},\ \bibinfo {pages} {191101} (\bibinfo {year} {2021})}\BibitemShut {NoStop}%
\bibitem [{\citenamefont {Schraivogel}\ \emph {et~al.}(2023)\citenamefont {Schraivogel}, \citenamefont {Christlmaier}, \citenamefont {López~Ríos}, \citenamefont {Alavi},\ and\ \citenamefont {Kats}}]{Schraivogel2023}%
  \BibitemOpen
  \bibfield  {author} {\bibinfo {author} {\bibfnamefont {T.}~\bibnamefont {Schraivogel}}, \bibinfo {author} {\bibfnamefont {E.~M.~C.}\ \bibnamefont {Christlmaier}}, \bibinfo {author} {\bibfnamefont {P.}~\bibnamefont {López~Ríos}}, \bibinfo {author} {\bibfnamefont {A.}~\bibnamefont {Alavi}}, \ and\ \bibinfo {author} {\bibfnamefont {D.}~\bibnamefont {Kats}},\ }\bibfield  {title} {\enquote {\bibinfo {title} {{Transcorrelated coupled cluster methods. II. Molecular systems}},}\ }\href {\doibase 10.1063/5.0151412} {\bibfield  {journal} {\bibinfo  {journal} {J. Chem. Phys.}\ }\textbf {\bibinfo {volume} {158}},\ \bibinfo {pages} {214106} (\bibinfo {year} {2023})}\BibitemShut {NoStop}%
\bibitem [{\citenamefont {Baiardi}, \citenamefont {Lesiuk},\ and\ \citenamefont {Reiher}(2022)}]{Baiardi2022}%
  \BibitemOpen
  \bibfield  {author} {\bibinfo {author} {\bibfnamefont {A.}~\bibnamefont {Baiardi}}, \bibinfo {author} {\bibfnamefont {M.}~\bibnamefont {Lesiuk}}, \ and\ \bibinfo {author} {\bibfnamefont {M.}~\bibnamefont {Reiher}},\ }\bibfield  {title} {\enquote {\bibinfo {title} {Explicitly correlated electronic structure calculations with transcorrelated matrix product operators},}\ }\href {\doibase 10.1021/acs.jctc.2c00167} {\bibfield  {journal} {\bibinfo  {journal} {J. Chem. Theory Comput.}\ }\textbf {\bibinfo {volume} {18}},\ \bibinfo {pages} {4203--4217} (\bibinfo {year} {2022})}\BibitemShut {NoStop}%
\bibitem [{\citenamefont {Liao}\ \emph {et~al.}(2023)\citenamefont {Liao}, \citenamefont {Zhai}, \citenamefont {Christlmaier}, \citenamefont {Schraivogel}, \citenamefont {Ríos}, \citenamefont {Kats},\ and\ \citenamefont {Alavi}}]{Liao2023}%
  \BibitemOpen
  \bibfield  {author} {\bibinfo {author} {\bibfnamefont {K.}~\bibnamefont {Liao}}, \bibinfo {author} {\bibfnamefont {H.}~\bibnamefont {Zhai}}, \bibinfo {author} {\bibfnamefont {E.~M.~C.}\ \bibnamefont {Christlmaier}}, \bibinfo {author} {\bibfnamefont {T.}~\bibnamefont {Schraivogel}}, \bibinfo {author} {\bibfnamefont {P.~L.}\ \bibnamefont {Ríos}}, \bibinfo {author} {\bibfnamefont {D.}~\bibnamefont {Kats}}, \ and\ \bibinfo {author} {\bibfnamefont {A.}~\bibnamefont {Alavi}},\ }\bibfield  {title} {\enquote {\bibinfo {title} {Density matrix renormalization group for transcorrelated hamiltonians: Ground and excited states in molecules},}\ }\href {\doibase 10.1021/acs.jctc.2c01207} {\bibfield  {journal} {\bibinfo  {journal} {J. Chem. Theory Comput.}\ }\textbf {\bibinfo {volume} {19}},\ \bibinfo {pages} {1734--1743} (\bibinfo {year} {2023})}\BibitemShut {NoStop}%
\bibitem [{\citenamefont {Guther}\ \emph {et~al.}(2021)\citenamefont {Guther}, \citenamefont {Cohen}, \citenamefont {Luo},\ and\ \citenamefont {Alavi}}]{Gunther2021}%
  \BibitemOpen
  \bibfield  {author} {\bibinfo {author} {\bibfnamefont {K.}~\bibnamefont {Guther}}, \bibinfo {author} {\bibfnamefont {A.~J.}\ \bibnamefont {Cohen}}, \bibinfo {author} {\bibfnamefont {H.}~\bibnamefont {Luo}}, \ and\ \bibinfo {author} {\bibfnamefont {A.}~\bibnamefont {Alavi}},\ }\bibfield  {title} {\enquote {\bibinfo {title} {{Binding curve of the beryllium dimer using similarity-transformed FCIQMC: Spectroscopic accuracy with triple-zeta basis sets}},}\ }\href {\doibase 10.1063/5.0055575} {\bibfield  {journal} {\bibinfo  {journal} {J. Chem. Phys.}\ }\textbf {\bibinfo {volume} {155}},\ \bibinfo {pages} {011102} (\bibinfo {year} {2021})}\BibitemShut {NoStop}%
\bibitem [{\citenamefont {Haupt}\ \emph {et~al.}(2023)\citenamefont {Haupt}, \citenamefont {Hosseini}, \citenamefont {López~Ríos}, \citenamefont {Dobrautz}, \citenamefont {Cohen},\ and\ \citenamefont {Alavi}}]{Haupt2023}%
  \BibitemOpen
  \bibfield  {author} {\bibinfo {author} {\bibfnamefont {J.~P.}\ \bibnamefont {Haupt}}, \bibinfo {author} {\bibfnamefont {S.~M.}\ \bibnamefont {Hosseini}}, \bibinfo {author} {\bibfnamefont {P.}~\bibnamefont {López~Ríos}}, \bibinfo {author} {\bibfnamefont {W.}~\bibnamefont {Dobrautz}}, \bibinfo {author} {\bibfnamefont {A.}~\bibnamefont {Cohen}}, \ and\ \bibinfo {author} {\bibfnamefont {A.}~\bibnamefont {Alavi}},\ }\bibfield  {title} {\enquote {\bibinfo {title} {{Optimizing Jastrow factors for the transcorrelated method}},}\ }\href {\doibase 10.1063/5.0147877} {\bibfield  {journal} {\bibinfo  {journal} {J. Chem. Phys.}\ }\textbf {\bibinfo {volume} {158}},\ \bibinfo {pages} {224105} (\bibinfo {year} {2023})}\BibitemShut {NoStop}%
\bibitem [{\citenamefont {Christlmaier}\ \emph {et~al.}(2023)\citenamefont {Christlmaier}, \citenamefont {Schraivogel}, \citenamefont {López~Ríos}, \citenamefont {Alavi},\ and\ \citenamefont {Kats}}]{Christlmaier2023}%
  \BibitemOpen
  \bibfield  {author} {\bibinfo {author} {\bibfnamefont {E.~M.~C.}\ \bibnamefont {Christlmaier}}, \bibinfo {author} {\bibfnamefont {T.}~\bibnamefont {Schraivogel}}, \bibinfo {author} {\bibfnamefont {P.}~\bibnamefont {López~Ríos}}, \bibinfo {author} {\bibfnamefont {A.}~\bibnamefont {Alavi}}, \ and\ \bibinfo {author} {\bibfnamefont {D.}~\bibnamefont {Kats}},\ }\bibfield  {title} {\enquote {\bibinfo {title} {{xTC: An efficient treatment of three-body interactions in transcorrelated methods}},}\ }\href {\doibase 10.1063/5.0154445} {\bibfield  {journal} {\bibinfo  {journal} {J. Chem. Phys.}\ }\textbf {\bibinfo {volume} {159}},\ \bibinfo {pages} {014113} (\bibinfo {year} {2023})}\BibitemShut {NoStop}%
\bibitem [{\citenamefont {Werner}, \citenamefont {Adler},\ and\ \citenamefont {Manby}(2007)}]{Werner2007}%
  \BibitemOpen
  \bibfield  {author} {\bibinfo {author} {\bibfnamefont {H.-J.}\ \bibnamefont {Werner}}, \bibinfo {author} {\bibfnamefont {T.~B.}\ \bibnamefont {Adler}}, \ and\ \bibinfo {author} {\bibfnamefont {F.~R.}\ \bibnamefont {Manby}},\ }\bibfield  {title} {\enquote {\bibinfo {title} {{General orbital invariant MP2-F12 theory}},}\ }\href {\doibase 10.1063/1.2712434} {\bibfield  {journal} {\bibinfo  {journal} {J. Chem. Phys.}\ }\textbf {\bibinfo {volume} {126}},\ \bibinfo {pages} {164102} (\bibinfo {year} {2007})}\BibitemShut {NoStop}%
\bibitem [{\citenamefont {Kats}\ \emph {et~al.}(2015)\citenamefont {Kats}, \citenamefont {Kreplin}, \citenamefont {Werner},\ and\ \citenamefont {Manby}}]{Kats2015}%
  \BibitemOpen
  \bibfield  {author} {\bibinfo {author} {\bibfnamefont {D.}~\bibnamefont {Kats}}, \bibinfo {author} {\bibfnamefont {D.}~\bibnamefont {Kreplin}}, \bibinfo {author} {\bibfnamefont {H.-J.}\ \bibnamefont {Werner}}, \ and\ \bibinfo {author} {\bibfnamefont {F.~R.}\ \bibnamefont {Manby}},\ }\bibfield  {title} {\enquote {\bibinfo {title} {{Accurate thermochemistry from explicitly correlated distinguishable cluster approximation}},}\ }\href {\doibase 10.1063/1.4907591} {\bibfield  {journal} {\bibinfo  {journal} {J. Chem. Phys.}\ }\textbf {\bibinfo {volume} {142}},\ \bibinfo {pages} {064111} (\bibinfo {year} {2015})}\BibitemShut {NoStop}%
\bibitem [{\citenamefont {Chakravorty}\ \emph {et~al.}(1993)\citenamefont {Chakravorty}, \citenamefont {Gwaltney}, \citenamefont {Davidson}, \citenamefont {Parpia},\ and\ \citenamefont {p~Fischer}}]{Chakravoty1993}%
  \BibitemOpen
  \bibfield  {author} {\bibinfo {author} {\bibfnamefont {S.~J.}\ \bibnamefont {Chakravorty}}, \bibinfo {author} {\bibfnamefont {S.~R.}\ \bibnamefont {Gwaltney}}, \bibinfo {author} {\bibfnamefont {E.~R.}\ \bibnamefont {Davidson}}, \bibinfo {author} {\bibfnamefont {F.~A.}\ \bibnamefont {Parpia}}, \ and\ \bibinfo {author} {\bibfnamefont {C.~F.}\ \bibnamefont {p~Fischer}},\ }\bibfield  {title} {\enquote {\bibinfo {title} {Ground-state correlation energies for atomic ions with 3 to 18 electrons},}\ }\href {\doibase 10.1103/PhysRevA.47.3649} {\bibfield  {journal} {\bibinfo  {journal} {Phys. Rev. A}\ }\textbf {\bibinfo {volume} {47}},\ \bibinfo {pages} {3649--3670} (\bibinfo {year} {1993})}\BibitemShut {NoStop}%
\bibitem [{\citenamefont {Drummond}, \citenamefont {Towler},\ and\ \citenamefont {Needs}(2004)}]{Drummond2004}%
  \BibitemOpen
  \bibfield  {author} {\bibinfo {author} {\bibfnamefont {N.~D.}\ \bibnamefont {Drummond}}, \bibinfo {author} {\bibfnamefont {M.~D.}\ \bibnamefont {Towler}}, \ and\ \bibinfo {author} {\bibfnamefont {R.~J.}\ \bibnamefont {Needs}},\ }\bibfield  {title} {\enquote {\bibinfo {title} {Jastrow correlation factor for atoms, molecules, and solids},}\ }\href {\doibase 10.1103/PhysRevB.70.235119} {\bibfield  {journal} {\bibinfo  {journal} {Phys. Rev. B}\ }\textbf {\bibinfo {volume} {70}},\ \bibinfo {pages} {235119} (\bibinfo {year} {2004})}\BibitemShut {NoStop}%
\bibitem [{\citenamefont {López~Ríos}\ \emph {et~al.}(2012)\citenamefont {López~Ríos}, \citenamefont {Seth}, \citenamefont {Drummond},\ and\ \citenamefont {Needs}}]{LopezRios2012}%
  \BibitemOpen
  \bibfield  {author} {\bibinfo {author} {\bibfnamefont {P.}~\bibnamefont {López~Ríos}}, \bibinfo {author} {\bibfnamefont {P.}~\bibnamefont {Seth}}, \bibinfo {author} {\bibfnamefont {N.~D.}\ \bibnamefont {Drummond}}, \ and\ \bibinfo {author} {\bibfnamefont {R.~J.}\ \bibnamefont {Needs}},\ }\bibfield  {title} {\enquote {\bibinfo {title} {Framework for constructing generic jastrow correlation factors},}\ }\href {\doibase https://doi.org/10.1103/PhysRevE.86.036703} {\bibfield  {journal} {\bibinfo  {journal} {Phys. Rev. E}\ }\textbf {\bibinfo {volume} {86}},\ \bibinfo {pages} {036703} (\bibinfo {year} {2012})}\BibitemShut {NoStop}%
\bibitem [{\citenamefont {Ma}\ \emph {et~al.}(2005)\citenamefont {Ma}, \citenamefont {Towler}, \citenamefont {Drummond},\ and\ \citenamefont {Needs}}]{Ma2005}%
  \BibitemOpen
  \bibfield  {author} {\bibinfo {author} {\bibfnamefont {A.}~\bibnamefont {Ma}}, \bibinfo {author} {\bibfnamefont {M.~D.}\ \bibnamefont {Towler}}, \bibinfo {author} {\bibfnamefont {N.~D.}\ \bibnamefont {Drummond}}, \ and\ \bibinfo {author} {\bibfnamefont {R.~J.}\ \bibnamefont {Needs}},\ }\bibfield  {title} {\enquote {\bibinfo {title} {{Scheme for adding electron–nucleus cusps to Gaussian orbitals}},}\ }\href {\doibase 10.1063/1.1940588} {\bibfield  {journal} {\bibinfo  {journal} {J. Chem. Phys.}\ }\textbf {\bibinfo {volume} {122}},\ \bibinfo {pages} {224322} (\bibinfo {year} {2005})}\BibitemShut {NoStop}%
\bibitem [{\citenamefont {Cleland}, \citenamefont {Booth},\ and\ \citenamefont {Alavi}(2010)}]{Cleland2010}%
  \BibitemOpen
  \bibfield  {author} {\bibinfo {author} {\bibfnamefont {D.}~\bibnamefont {Cleland}}, \bibinfo {author} {\bibfnamefont {G.~H.}\ \bibnamefont {Booth}}, \ and\ \bibinfo {author} {\bibfnamefont {A.}~\bibnamefont {Alavi}},\ }\bibfield  {title} {\enquote {\bibinfo {title} {Communications: Survival of the fittest: Accelerating convergence in full configuration-interaction quantum monte carlo},}\ }\href {\doibase 10.1063/1.3302277} {\bibfield  {journal} {\bibinfo  {journal} {J. Chem. Phys.}\ }\textbf {\bibinfo {volume} {132}} (\bibinfo {year} {2010}),\ 10.1063/1.3302277}\BibitemShut {NoStop}%
\bibitem [{\citenamefont {Woon}\ and\ \citenamefont {Dunning}(1993)}]{Woon1993}%
  \BibitemOpen
  \bibfield  {author} {\bibinfo {author} {\bibfnamefont {D.~E.}\ \bibnamefont {Woon}}\ and\ \bibinfo {author} {\bibfnamefont {T.~H.}\ \bibnamefont {Dunning}},\ }\bibfield  {title} {\enquote {\bibinfo {title} {Gaussian basis sets for use in correlated molecular calculations. iii. the atoms aluminum through argon},}\ }\href {\doibase 10.1063/1.464303} {\bibfield  {journal} {\bibinfo  {journal} {J. Chem. Phys.}\ }\textbf {\bibinfo {volume} {98}},\ \bibinfo {pages} {1358--1371} (\bibinfo {year} {1993})}\BibitemShut {NoStop}%
\bibitem [{\citenamefont {Peterson}\ and\ \citenamefont {Dunning}(2002)}]{Peterson2002}%
  \BibitemOpen
  \bibfield  {author} {\bibinfo {author} {\bibfnamefont {K.~A.}\ \bibnamefont {Peterson}}\ and\ \bibinfo {author} {\bibfnamefont {T.~H.}\ \bibnamefont {Dunning}},\ }\bibfield  {title} {\enquote {\bibinfo {title} {Accurate correlation consistent basis sets for molecular core-valence correlation effects: The second row atoms al-ar, and the first row atoms b-ne revisited},}\ }\href {\doibase 10.1063/1.1520138} {\bibfield  {journal} {\bibinfo  {journal} {J. Chem. Phys.}\ }\textbf {\bibinfo {volume} {117}},\ \bibinfo {pages} {10548--10560} (\bibinfo {year} {2002})}\BibitemShut {NoStop}%
\bibitem [{\citenamefont {Prascher}\ \emph {et~al.}(2011)\citenamefont {Prascher}, \citenamefont {Woon}, \citenamefont {Peterson}, \citenamefont {Dunning},\ and\ \citenamefont {Wilson}}]{Prascher2011}%
  \BibitemOpen
  \bibfield  {author} {\bibinfo {author} {\bibfnamefont {B.~P.}\ \bibnamefont {Prascher}}, \bibinfo {author} {\bibfnamefont {D.~E.}\ \bibnamefont {Woon}}, \bibinfo {author} {\bibfnamefont {K.~A.}\ \bibnamefont {Peterson}}, \bibinfo {author} {\bibfnamefont {T.~H.}\ \bibnamefont {Dunning}}, \ and\ \bibinfo {author} {\bibfnamefont {A.~K.}\ \bibnamefont {Wilson}},\ }\bibfield  {title} {\enquote {\bibinfo {title} {Gaussian basis sets for use in correlated molecular calculations. vii. valence, core-valence, and scalar relativistic basis sets for li, be, na, and mg},}\ }\href {\doibase 10.1007/s00214-010-0764-0} {\bibfield  {journal} {\bibinfo  {journal} {Theor. Chem. Acc.}\ }\textbf {\bibinfo {volume} {128}},\ \bibinfo {pages} {69--82} (\bibinfo {year} {2011})}\BibitemShut {NoStop}%
\bibitem [{\citenamefont {Sun}(2015)}]{Sun2015}%
  \BibitemOpen
  \bibfield  {author} {\bibinfo {author} {\bibfnamefont {Q.}~\bibnamefont {Sun}},\ }\bibfield  {title} {\enquote {\bibinfo {title} {Libcint: An efficient general integral library for gaussian basis functions},}\ }\href {\doibase https://doi.org/10.1002/jcc.23981} {\bibfield  {journal} {\bibinfo  {journal} {J. Comput. Chem.}\ }\textbf {\bibinfo {volume} {36}},\ \bibinfo {pages} {1664--1671} (\bibinfo {year} {2015})}\BibitemShut {NoStop}%
\bibitem [{\citenamefont {Sun}\ \emph {et~al.}(2018)\citenamefont {Sun}, \citenamefont {Berkelbach}, \citenamefont {Blunt}, \citenamefont {Booth}, \citenamefont {Guo}, \citenamefont {Li}, \citenamefont {Liu}, \citenamefont {McClain}, \citenamefont {Sayfutyarova}, \citenamefont {Sharma}, \citenamefont {Wouters},\ and\ \citenamefont {Chan}}]{Sun2018}%
  \BibitemOpen
  \bibfield  {author} {\bibinfo {author} {\bibfnamefont {Q.}~\bibnamefont {Sun}}, \bibinfo {author} {\bibfnamefont {T.~C.}\ \bibnamefont {Berkelbach}}, \bibinfo {author} {\bibfnamefont {N.~S.}\ \bibnamefont {Blunt}}, \bibinfo {author} {\bibfnamefont {G.~H.}\ \bibnamefont {Booth}}, \bibinfo {author} {\bibfnamefont {S.}~\bibnamefont {Guo}}, \bibinfo {author} {\bibfnamefont {Z.}~\bibnamefont {Li}}, \bibinfo {author} {\bibfnamefont {J.}~\bibnamefont {Liu}}, \bibinfo {author} {\bibfnamefont {J.~D.}\ \bibnamefont {McClain}}, \bibinfo {author} {\bibfnamefont {E.~R.}\ \bibnamefont {Sayfutyarova}}, \bibinfo {author} {\bibfnamefont {S.}~\bibnamefont {Sharma}}, \bibinfo {author} {\bibfnamefont {S.}~\bibnamefont {Wouters}}, \ and\ \bibinfo {author} {\bibfnamefont {G.~K.-L.}\ \bibnamefont {Chan}},\ }\bibfield  {title} {\enquote {\bibinfo {title} {Pyscf: the python-based simulations of chemistry framework},}\ }\href {\doibase https://doi.org/10.1002/wcms.1340} {\bibfield  {journal} {\bibinfo  {journal} {WIREs Comput. Mol.
  Sci.}\ }\textbf {\bibinfo {volume} {8}},\ \bibinfo {pages} {e1340} (\bibinfo {year} {2018})}\BibitemShut {NoStop}%
\bibitem [{\citenamefont {Sun}\ \emph {et~al.}(2020)\citenamefont {Sun}, \citenamefont {Zhang}, \citenamefont {Banerjee}, \citenamefont {Bao}, \citenamefont {Barbry}, \citenamefont {Blunt}, \citenamefont {Bogdanov}, \citenamefont {Booth}, \citenamefont {Chen}, \citenamefont {Cui}, \citenamefont {Eriksen}, \citenamefont {Gao}, \citenamefont {Guo}, \citenamefont {Hermann}, \citenamefont {Hermes}, \citenamefont {Koh}, \citenamefont {Koval}, \citenamefont {Lehtola}, \citenamefont {Li}, \citenamefont {Liu}, \citenamefont {Mardirossian}, \citenamefont {McClain}, \citenamefont {Motta}, \citenamefont {Mussard}, \citenamefont {Pham}, \citenamefont {Pulkin}, \citenamefont {Purwanto}, \citenamefont {Robinson}, \citenamefont {Ronca}, \citenamefont {Sayfutyarova}, \citenamefont {Scheurer}, \citenamefont {Schurkus}, \citenamefont {Smith}, \citenamefont {Sun}, \citenamefont {Sun}, \citenamefont {Upadhyay}, \citenamefont {Wagner}, \citenamefont {Wang}, \citenamefont {White}, \citenamefont {Whitfield}, \citenamefont
  {Williamson}, \citenamefont {Wouters}, \citenamefont {Yang}, \citenamefont {Yu}, \citenamefont {Zhu}, \citenamefont {Berkelbach}, \citenamefont {Sharma}, \citenamefont {Sokolov},\ and\ \citenamefont {Chan}}]{Sun2020}%
  \BibitemOpen
  \bibfield  {author} {\bibinfo {author} {\bibfnamefont {Q.}~\bibnamefont {Sun}}, \bibinfo {author} {\bibfnamefont {X.}~\bibnamefont {Zhang}}, \bibinfo {author} {\bibfnamefont {S.}~\bibnamefont {Banerjee}}, \bibinfo {author} {\bibfnamefont {P.}~\bibnamefont {Bao}}, \bibinfo {author} {\bibfnamefont {M.}~\bibnamefont {Barbry}}, \bibinfo {author} {\bibfnamefont {N.~S.}\ \bibnamefont {Blunt}}, \bibinfo {author} {\bibfnamefont {N.~A.}\ \bibnamefont {Bogdanov}}, \bibinfo {author} {\bibfnamefont {G.~H.}\ \bibnamefont {Booth}}, \bibinfo {author} {\bibfnamefont {J.}~\bibnamefont {Chen}}, \bibinfo {author} {\bibfnamefont {Z.-H.}\ \bibnamefont {Cui}}, \bibinfo {author} {\bibfnamefont {J.~J.}\ \bibnamefont {Eriksen}}, \bibinfo {author} {\bibfnamefont {Y.}~\bibnamefont {Gao}}, \bibinfo {author} {\bibfnamefont {S.}~\bibnamefont {Guo}}, \bibinfo {author} {\bibfnamefont {J.}~\bibnamefont {Hermann}}, \bibinfo {author} {\bibfnamefont {M.~R.}\ \bibnamefont {Hermes}}, \bibinfo {author} {\bibfnamefont {K.}~\bibnamefont {Koh}},
  \bibinfo {author} {\bibfnamefont {P.}~\bibnamefont {Koval}}, \bibinfo {author} {\bibfnamefont {S.}~\bibnamefont {Lehtola}}, \bibinfo {author} {\bibfnamefont {Z.}~\bibnamefont {Li}}, \bibinfo {author} {\bibfnamefont {J.}~\bibnamefont {Liu}}, \bibinfo {author} {\bibfnamefont {N.}~\bibnamefont {Mardirossian}}, \bibinfo {author} {\bibfnamefont {J.~D.}\ \bibnamefont {McClain}}, \bibinfo {author} {\bibfnamefont {M.}~\bibnamefont {Motta}}, \bibinfo {author} {\bibfnamefont {B.}~\bibnamefont {Mussard}}, \bibinfo {author} {\bibfnamefont {H.~Q.}\ \bibnamefont {Pham}}, \bibinfo {author} {\bibfnamefont {A.}~\bibnamefont {Pulkin}}, \bibinfo {author} {\bibfnamefont {W.}~\bibnamefont {Purwanto}}, \bibinfo {author} {\bibfnamefont {P.~J.}\ \bibnamefont {Robinson}}, \bibinfo {author} {\bibfnamefont {E.}~\bibnamefont {Ronca}}, \bibinfo {author} {\bibfnamefont {E.~R.}\ \bibnamefont {Sayfutyarova}}, \bibinfo {author} {\bibfnamefont {M.}~\bibnamefont {Scheurer}}, \bibinfo {author} {\bibfnamefont {H.~F.}\ \bibnamefont {Schurkus}},
  \bibinfo {author} {\bibfnamefont {J.~E.~T.}\ \bibnamefont {Smith}}, \bibinfo {author} {\bibfnamefont {C.}~\bibnamefont {Sun}}, \bibinfo {author} {\bibfnamefont {S.-N.}\ \bibnamefont {Sun}}, \bibinfo {author} {\bibfnamefont {S.}~\bibnamefont {Upadhyay}}, \bibinfo {author} {\bibfnamefont {L.~K.}\ \bibnamefont {Wagner}}, \bibinfo {author} {\bibfnamefont {X.}~\bibnamefont {Wang}}, \bibinfo {author} {\bibfnamefont {A.}~\bibnamefont {White}}, \bibinfo {author} {\bibfnamefont {J.~D.}\ \bibnamefont {Whitfield}}, \bibinfo {author} {\bibfnamefont {M.~J.}\ \bibnamefont {Williamson}}, \bibinfo {author} {\bibfnamefont {S.}~\bibnamefont {Wouters}}, \bibinfo {author} {\bibfnamefont {J.}~\bibnamefont {Yang}}, \bibinfo {author} {\bibfnamefont {J.~M.}\ \bibnamefont {Yu}}, \bibinfo {author} {\bibfnamefont {T.}~\bibnamefont {Zhu}}, \bibinfo {author} {\bibfnamefont {T.~C.}\ \bibnamefont {Berkelbach}}, \bibinfo {author} {\bibfnamefont {S.}~\bibnamefont {Sharma}}, \bibinfo {author} {\bibfnamefont {A.~Y.}\ \bibnamefont {Sokolov}},
  \ and\ \bibinfo {author} {\bibfnamefont {G.~K.-L.}\ \bibnamefont {Chan}},\ }\bibfield  {title} {\enquote {\bibinfo {title} {{Recent developments in the PySCF program package}},}\ }\href {\doibase 10.1063/5.0006074} {\bibfield  {journal} {\bibinfo  {journal} {J. Chem. Phys.}\ }\textbf {\bibinfo {volume} {153}},\ \bibinfo {pages} {024109} (\bibinfo {year} {2020})}\BibitemShut {NoStop}%
\bibitem [{\citenamefont {Needs}\ \emph {et~al.}(2020)\citenamefont {Needs}, \citenamefont {Towler}, \citenamefont {Drummond}, \citenamefont {López~Ríos},\ and\ \citenamefont {Trail}}]{Needs2020}%
  \BibitemOpen
  \bibfield  {author} {\bibinfo {author} {\bibfnamefont {R.~J.}\ \bibnamefont {Needs}}, \bibinfo {author} {\bibfnamefont {M.~D.}\ \bibnamefont {Towler}}, \bibinfo {author} {\bibfnamefont {N.~D.}\ \bibnamefont {Drummond}}, \bibinfo {author} {\bibfnamefont {P.}~\bibnamefont {López~Ríos}}, \ and\ \bibinfo {author} {\bibfnamefont {J.~R.}\ \bibnamefont {Trail}},\ }\bibfield  {title} {\enquote {\bibinfo {title} {{Variational and diffusion quantum Monte Carlo calculations with the CASINO code}},}\ }\href {\doibase 10.1063/1.5144288} {\bibfield  {journal} {\bibinfo  {journal} {J. Chem. Phys.}\ }\textbf {\bibinfo {volume} {152}},\ \bibinfo {pages} {154106} (\bibinfo {year} {2020})}\BibitemShut {NoStop}%
\bibitem [{tch()}]{tchint}%
  \BibitemOpen
  \href@noop {} {\ }\bibinfo {note} {Transcorrelated Hamiltonian integral library tchint to be released; available from the authors upon reasonable request}\BibitemShut {NoStop}%
\bibitem [{\citenamefont {Guther}\ \emph {et~al.}(2020)\citenamefont {Guther}, \citenamefont {Anderson}, \citenamefont {Blunt}, \citenamefont {Bogdanov}, \citenamefont {Cleland}, \citenamefont {Dattani}, \citenamefont {Dobrautz}, \citenamefont {Ghanem}, \citenamefont {Jeszenszki}, \citenamefont {Liebermann}, \citenamefont {Manni}, \citenamefont {Lozovoi}, \citenamefont {Luo}, \citenamefont {Ma}, \citenamefont {Merz}, \citenamefont {Overy}, \citenamefont {Rampp}, \citenamefont {Samanta}, \citenamefont {Schwarz}, \citenamefont {Shepherd}, \citenamefont {Smart}, \citenamefont {Vitale}, \citenamefont {Weser}, \citenamefont {Booth},\ and\ \citenamefont {Alavi}}]{Kai2020}%
  \BibitemOpen
  \bibfield  {author} {\bibinfo {author} {\bibfnamefont {K.}~\bibnamefont {Guther}}, \bibinfo {author} {\bibfnamefont {R.~J.}\ \bibnamefont {Anderson}}, \bibinfo {author} {\bibfnamefont {N.~S.}\ \bibnamefont {Blunt}}, \bibinfo {author} {\bibfnamefont {N.~A.}\ \bibnamefont {Bogdanov}}, \bibinfo {author} {\bibfnamefont {D.}~\bibnamefont {Cleland}}, \bibinfo {author} {\bibfnamefont {N.}~\bibnamefont {Dattani}}, \bibinfo {author} {\bibfnamefont {W.}~\bibnamefont {Dobrautz}}, \bibinfo {author} {\bibfnamefont {K.}~\bibnamefont {Ghanem}}, \bibinfo {author} {\bibfnamefont {P.}~\bibnamefont {Jeszenszki}}, \bibinfo {author} {\bibfnamefont {N.}~\bibnamefont {Liebermann}}, \bibinfo {author} {\bibfnamefont {G.~L.}\ \bibnamefont {Manni}}, \bibinfo {author} {\bibfnamefont {A.~Y.}\ \bibnamefont {Lozovoi}}, \bibinfo {author} {\bibfnamefont {H.}~\bibnamefont {Luo}}, \bibinfo {author} {\bibfnamefont {D.}~\bibnamefont {Ma}}, \bibinfo {author} {\bibfnamefont {F.}~\bibnamefont {Merz}}, \bibinfo {author} {\bibfnamefont
  {C.}~\bibnamefont {Overy}}, \bibinfo {author} {\bibfnamefont {M.}~\bibnamefont {Rampp}}, \bibinfo {author} {\bibfnamefont {P.~K.}\ \bibnamefont {Samanta}}, \bibinfo {author} {\bibfnamefont {L.~R.}\ \bibnamefont {Schwarz}}, \bibinfo {author} {\bibfnamefont {J.~J.}\ \bibnamefont {Shepherd}}, \bibinfo {author} {\bibfnamefont {S.~D.}\ \bibnamefont {Smart}}, \bibinfo {author} {\bibfnamefont {E.}~\bibnamefont {Vitale}}, \bibinfo {author} {\bibfnamefont {O.}~\bibnamefont {Weser}}, \bibinfo {author} {\bibfnamefont {G.~H.}\ \bibnamefont {Booth}}, \ and\ \bibinfo {author} {\bibfnamefont {A.}~\bibnamefont {Alavi}},\ }\bibfield  {title} {\enquote {\bibinfo {title} {{NECI: N-Electron Configuration Interaction with an emphasis on state-of-the-art stochastic methods}},}\ }\href {\doibase 10.1063/5.0005754} {\bibfield  {journal} {\bibinfo  {journal} {J. Chem. Phys.}\ }\textbf {\bibinfo {volume} {153}},\ \bibinfo {pages} {034107} (\bibinfo {year} {2020})}\BibitemShut {NoStop}%
\bibitem [{\citenamefont {Werner}\ \emph {et~al.}(2012)\citenamefont {Werner}, \citenamefont {Knowles}, \citenamefont {Knizia}, \citenamefont {Manby},\ and\ \citenamefont {Sch{\"u}tz}}]{MOLPRO-WIREs}%
  \BibitemOpen
  \bibfield  {author} {\bibinfo {author} {\bibfnamefont {H.-J.}\ \bibnamefont {Werner}}, \bibinfo {author} {\bibfnamefont {P.~J.}\ \bibnamefont {Knowles}}, \bibinfo {author} {\bibfnamefont {G.}~\bibnamefont {Knizia}}, \bibinfo {author} {\bibfnamefont {F.~R.}\ \bibnamefont {Manby}}, \ and\ \bibinfo {author} {\bibfnamefont {M.}~\bibnamefont {Sch{\"u}tz}},\ }\bibfield  {title} {\enquote {\bibinfo {title} {{Molpro: a general-purpose quantum chemistry program package}},}\ }\href@noop {} {\bibfield  {journal} {\bibinfo  {journal} {WIREs Comput Mol Sci}\ }\textbf {\bibinfo {volume} {2}},\ \bibinfo {pages} {242--253} (\bibinfo {year} {2012})}\BibitemShut {NoStop}%
\bibitem [{\citenamefont {Werner}\ \emph {et~al.}(2020)\citenamefont {Werner}, \citenamefont {Knowles}, \citenamefont {Manby}, \citenamefont {Black}, \citenamefont {Doll}, \citenamefont {Heßelmann}, \citenamefont {Kats}, \citenamefont {Köhn}, \citenamefont {Korona}, \citenamefont {Kreplin}, \citenamefont {Ma}, \citenamefont {Miller}, \citenamefont {Mitrushchenkov}, \citenamefont {Peterson}, \citenamefont {Polyak}, \citenamefont {Rauhut},\ and\ \citenamefont {Sibaev}}]{MolproJCP}%
  \BibitemOpen
  \bibfield  {author} {\bibinfo {author} {\bibfnamefont {H.-J.}\ \bibnamefont {Werner}}, \bibinfo {author} {\bibfnamefont {P.~J.}\ \bibnamefont {Knowles}}, \bibinfo {author} {\bibfnamefont {F.~R.}\ \bibnamefont {Manby}}, \bibinfo {author} {\bibfnamefont {J.~A.}\ \bibnamefont {Black}}, \bibinfo {author} {\bibfnamefont {K.}~\bibnamefont {Doll}}, \bibinfo {author} {\bibfnamefont {A.}~\bibnamefont {Heßelmann}}, \bibinfo {author} {\bibfnamefont {D.}~\bibnamefont {Kats}}, \bibinfo {author} {\bibfnamefont {A.}~\bibnamefont {Köhn}}, \bibinfo {author} {\bibfnamefont {T.}~\bibnamefont {Korona}}, \bibinfo {author} {\bibfnamefont {D.~A.}\ \bibnamefont {Kreplin}}, \bibinfo {author} {\bibfnamefont {Q.}~\bibnamefont {Ma}}, \bibinfo {author} {\bibfnamefont {I.}~\bibnamefont {Miller}, \bibfnamefont {Thomas~F.}}, \bibinfo {author} {\bibfnamefont {A.}~\bibnamefont {Mitrushchenkov}}, \bibinfo {author} {\bibfnamefont {K.~A.}\ \bibnamefont {Peterson}}, \bibinfo {author} {\bibfnamefont {I.}~\bibnamefont {Polyak}}, \bibinfo
  {author} {\bibfnamefont {G.}~\bibnamefont {Rauhut}}, \ and\ \bibinfo {author} {\bibfnamefont {M.}~\bibnamefont {Sibaev}},\ }\bibfield  {title} {\enquote {\bibinfo {title} {{The Molpro quantum chemistry package}},}\ }\href {\doibase 10.1063/5.0005081} {\bibfield  {journal} {\bibinfo  {journal} {J. Chem. Phys.}\ }\textbf {\bibinfo {volume} {152}},\ \bibinfo {pages} {144107} (\bibinfo {year} {2020})}\BibitemShut {NoStop}%
\bibitem [{\citenamefont {Werner}\ \emph {et~al.}()\citenamefont {Werner}, \citenamefont {Knowles}, \citenamefont {Celani}, \citenamefont {Gy\"orffy}, \citenamefont {Hesselmann}, \citenamefont {Kats}, \citenamefont {Knizia}, \citenamefont {K\"ohn}, \citenamefont {Korona}, \citenamefont {Kreplin}, \citenamefont {Lindh}, \citenamefont {Ma}, \citenamefont {Manby}, \citenamefont {Mitrushenkov}, \citenamefont {Rauhut}, \citenamefont {{Sch\"{u}tz}}, \citenamefont {Shamasundar}, \citenamefont {Adler}, \citenamefont {Amos}, \citenamefont {Bennie}, \citenamefont {Bernhardsson}, \citenamefont {Berning}, \citenamefont {Black}, \citenamefont {Bygrave}, \citenamefont {Cimiraglia}, \citenamefont {Cooper}, \citenamefont {Coughtrie}, \citenamefont {Deegan}, \citenamefont {Dobbyn}, \citenamefont {Doll}, \citenamefont {Dornbach}, \citenamefont {Eckert}, \citenamefont {Erfort}, \citenamefont {Goll}, \citenamefont {Hampel}, \citenamefont {Hetzer}, \citenamefont {Hill}, \citenamefont {Hodges}, \citenamefont {Hrenar},
  \citenamefont {Jansen}, \citenamefont {K\"oppl}, \citenamefont {Kollmar}, \citenamefont {Lee}, \citenamefont {Liu}, \citenamefont {Lloyd}, \citenamefont {Mata}, \citenamefont {May}, \citenamefont {Mussard}, \citenamefont {McNicholas}, \citenamefont {Meyer}, \citenamefont {{Miller III}}, \citenamefont {Mura}, \citenamefont {Nicklass}, \citenamefont {O'Neill}, \citenamefont {Palmieri}, \citenamefont {Peng}, \citenamefont {Peterson}, \citenamefont {Pfl\"uger}, \citenamefont {Pitzer}, \citenamefont {Polyak}, \citenamefont {Reiher}, \citenamefont {Richardson}, \citenamefont {Robinson}, \citenamefont {Schr\"oder}, \citenamefont {Schwilk}, \citenamefont {Shiozaki}, \citenamefont {Sibaev}, \citenamefont {Stoll}, \citenamefont {Stone}, \citenamefont {Tarroni}, \citenamefont {Thorsteinsson}, \citenamefont {Toulouse}, \citenamefont {Wang}, \citenamefont {Welborn},\ and\ \citenamefont {Ziegler}}]{MOLPRO}%
  \BibitemOpen
  \bibfield  {author} {\bibinfo {author} {\bibfnamefont {H.-J.}\ \bibnamefont {Werner}}, \bibinfo {author} {\bibfnamefont {P.~J.}\ \bibnamefont {Knowles}}, \bibinfo {author} {\bibfnamefont {P.}~\bibnamefont {Celani}}, \bibinfo {author} {\bibfnamefont {W.}~\bibnamefont {Gy\"orffy}}, \bibinfo {author} {\bibfnamefont {A.}~\bibnamefont {Hesselmann}}, \bibinfo {author} {\bibfnamefont {D.}~\bibnamefont {Kats}}, \bibinfo {author} {\bibfnamefont {G.}~\bibnamefont {Knizia}}, \bibinfo {author} {\bibfnamefont {A.}~\bibnamefont {K\"ohn}}, \bibinfo {author} {\bibfnamefont {T.}~\bibnamefont {Korona}}, \bibinfo {author} {\bibfnamefont {D.}~\bibnamefont {Kreplin}}, \bibinfo {author} {\bibfnamefont {R.}~\bibnamefont {Lindh}}, \bibinfo {author} {\bibfnamefont {Q.}~\bibnamefont {Ma}}, \bibinfo {author} {\bibfnamefont {F.~R.}\ \bibnamefont {Manby}}, \bibinfo {author} {\bibfnamefont {A.}~\bibnamefont {Mitrushenkov}}, \bibinfo {author} {\bibfnamefont {G.}~\bibnamefont {Rauhut}}, \bibinfo {author} {\bibfnamefont {M.}~\bibnamefont
  {{Sch\"{u}tz}}}, \bibinfo {author} {\bibfnamefont {K.~R.}\ \bibnamefont {Shamasundar}}, \bibinfo {author} {\bibfnamefont {T.~B.}\ \bibnamefont {Adler}}, \bibinfo {author} {\bibfnamefont {R.~D.}\ \bibnamefont {Amos}}, \bibinfo {author} {\bibfnamefont {S.~J.}\ \bibnamefont {Bennie}}, \bibinfo {author} {\bibfnamefont {A.}~\bibnamefont {Bernhardsson}}, \bibinfo {author} {\bibfnamefont {A.}~\bibnamefont {Berning}}, \bibinfo {author} {\bibfnamefont {J.~A.}\ \bibnamefont {Black}}, \bibinfo {author} {\bibfnamefont {P.~J.}\ \bibnamefont {Bygrave}}, \bibinfo {author} {\bibfnamefont {R.}~\bibnamefont {Cimiraglia}}, \bibinfo {author} {\bibfnamefont {D.~L.}\ \bibnamefont {Cooper}}, \bibinfo {author} {\bibfnamefont {D.}~\bibnamefont {Coughtrie}}, \bibinfo {author} {\bibfnamefont {M.~J.~O.}\ \bibnamefont {Deegan}}, \bibinfo {author} {\bibfnamefont {A.~J.}\ \bibnamefont {Dobbyn}}, \bibinfo {author} {\bibfnamefont {K.}~\bibnamefont {Doll}}, \bibinfo {author} {\bibfnamefont {M.}~\bibnamefont {Dornbach}}, \bibinfo {author}
  {\bibfnamefont {F.}~\bibnamefont {Eckert}}, \bibinfo {author} {\bibfnamefont {S.}~\bibnamefont {Erfort}}, \bibinfo {author} {\bibfnamefont {E.}~\bibnamefont {Goll}}, \bibinfo {author} {\bibfnamefont {C.}~\bibnamefont {Hampel}}, \bibinfo {author} {\bibfnamefont {G.}~\bibnamefont {Hetzer}}, \bibinfo {author} {\bibfnamefont {J.~G.}\ \bibnamefont {Hill}}, \bibinfo {author} {\bibfnamefont {M.}~\bibnamefont {Hodges}}, \bibinfo {author} {\bibfnamefont {T.}~\bibnamefont {Hrenar}}, \bibinfo {author} {\bibfnamefont {G.}~\bibnamefont {Jansen}}, \bibinfo {author} {\bibfnamefont {C.}~\bibnamefont {K\"oppl}}, \bibinfo {author} {\bibfnamefont {C.}~\bibnamefont {Kollmar}}, \bibinfo {author} {\bibfnamefont {S.~J.~R.}\ \bibnamefont {Lee}}, \bibinfo {author} {\bibfnamefont {Y.}~\bibnamefont {Liu}}, \bibinfo {author} {\bibfnamefont {A.~W.}\ \bibnamefont {Lloyd}}, \bibinfo {author} {\bibfnamefont {R.~A.}\ \bibnamefont {Mata}}, \bibinfo {author} {\bibfnamefont {A.~J.}\ \bibnamefont {May}}, \bibinfo {author} {\bibfnamefont
  {B.}~\bibnamefont {Mussard}}, \bibinfo {author} {\bibfnamefont {S.~J.}\ \bibnamefont {McNicholas}}, \bibinfo {author} {\bibfnamefont {W.}~\bibnamefont {Meyer}}, \bibinfo {author} {\bibfnamefont {T.~F.}\ \bibnamefont {{Miller III}}}, \bibinfo {author} {\bibfnamefont {M.~E.}\ \bibnamefont {Mura}}, \bibinfo {author} {\bibfnamefont {A.}~\bibnamefont {Nicklass}}, \bibinfo {author} {\bibfnamefont {D.~P.}\ \bibnamefont {O'Neill}}, \bibinfo {author} {\bibfnamefont {P.}~\bibnamefont {Palmieri}}, \bibinfo {author} {\bibfnamefont {D.}~\bibnamefont {Peng}}, \bibinfo {author} {\bibfnamefont {K.~A.}\ \bibnamefont {Peterson}}, \bibinfo {author} {\bibfnamefont {K.}~\bibnamefont {Pfl\"uger}}, \bibinfo {author} {\bibfnamefont {R.}~\bibnamefont {Pitzer}}, \bibinfo {author} {\bibfnamefont {I.}~\bibnamefont {Polyak}}, \bibinfo {author} {\bibfnamefont {M.}~\bibnamefont {Reiher}}, \bibinfo {author} {\bibfnamefont {J.~O.}\ \bibnamefont {Richardson}}, \bibinfo {author} {\bibfnamefont {J.~B.}\ \bibnamefont {Robinson}}, \bibinfo
  {author} {\bibfnamefont {B.}~\bibnamefont {Schr\"oder}}, \bibinfo {author} {\bibfnamefont {M.}~\bibnamefont {Schwilk}}, \bibinfo {author} {\bibfnamefont {T.}~\bibnamefont {Shiozaki}}, \bibinfo {author} {\bibfnamefont {M.}~\bibnamefont {Sibaev}}, \bibinfo {author} {\bibfnamefont {H.}~\bibnamefont {Stoll}}, \bibinfo {author} {\bibfnamefont {A.~J.}\ \bibnamefont {Stone}}, \bibinfo {author} {\bibfnamefont {R.}~\bibnamefont {Tarroni}}, \bibinfo {author} {\bibfnamefont {T.}~\bibnamefont {Thorsteinsson}}, \bibinfo {author} {\bibfnamefont {J.}~\bibnamefont {Toulouse}}, \bibinfo {author} {\bibfnamefont {M.}~\bibnamefont {Wang}}, \bibinfo {author} {\bibfnamefont {M.}~\bibnamefont {Welborn}}, \ and\ \bibinfo {author} {\bibfnamefont {B.}~\bibnamefont {Ziegler}},\ }\href@noop {} {\enquote {\bibinfo {title} {Molpro, version 2022.3.0, a package of ab initio programs},}\ }\bibinfo {note} {See https://www.molpro.net/}\BibitemShut {NoStop}%
\bibitem [{\citenamefont {Kats}\ \emph {et~al.}(2024{\natexlab{a}})\citenamefont {Kats}, \citenamefont {Schraivogel}, \citenamefont {Hauskrecht}, \citenamefont {Rickert},\ and\ \citenamefont {Wu}}]{elemcoil}%
  \BibitemOpen
  \bibfield  {author} {\bibinfo {author} {\bibfnamefont {D.}~\bibnamefont {Kats}}, \bibinfo {author} {\bibfnamefont {T.}~\bibnamefont {Schraivogel}}, \bibinfo {author} {\bibfnamefont {J.}~\bibnamefont {Hauskrecht}}, \bibinfo {author} {\bibfnamefont {C.}~\bibnamefont {Rickert}}, \ and\ \bibinfo {author} {\bibfnamefont {F.}~\bibnamefont {Wu}},\ }\href@noop {} {\enquote {\bibinfo {title} {\texttt{ElemCo.jl}: Julia program package for electron correlation methods},}\ } (\bibinfo {year} {2024}{\natexlab{a}}),\ \bibinfo {note} {see github.com/fkfest/ElemCo.jl}\BibitemShut {NoStop}%
\bibitem [{\citenamefont {Kats}\ \emph {et~al.}(2024{\natexlab{b}})\citenamefont {Kats}, \citenamefont {Christlmaier}, \citenamefont {Schraivogel},\ and\ \citenamefont {Alavi}}]{katsOrbital2024}%
  \BibitemOpen
  \bibfield  {author} {\bibinfo {author} {\bibfnamefont {D.}~\bibnamefont {Kats}}, \bibinfo {author} {\bibfnamefont {E.~M.~C.}\ \bibnamefont {Christlmaier}}, \bibinfo {author} {\bibfnamefont {T.}~\bibnamefont {Schraivogel}}, \ and\ \bibinfo {author} {\bibfnamefont {A.}~\bibnamefont {Alavi}},\ }\bibfield  {title} {\enquote {\bibinfo {title} {Orbital optimisation in {{xTC}} transcorrelated methods},}\ }\href {\doibase 10.1039/d4fd00036f} {\bibfield  {journal} {\bibinfo  {journal} {Faraday Discuss.}\ } (\bibinfo {year} {2024}{\natexlab{b}}),\ 10.1039/d4fd00036f}\BibitemShut {NoStop}%
\bibitem [{\citenamefont {Kats}\ and\ \citenamefont {Manby}(2013)}]{kats_dc_2013}%
  \BibitemOpen
  \bibfield  {author} {\bibinfo {author} {\bibfnamefont {D.}~\bibnamefont {Kats}}\ and\ \bibinfo {author} {\bibfnamefont {F.~R.}\ \bibnamefont {Manby}},\ }\bibfield  {title} {\enquote {\bibinfo {title} {Communication: The distinguishable cluster approximation},}\ }\href {\doibase doi:10.1063/1.4813481} {\bibfield  {journal} {\bibinfo  {journal} {J. Chem. Phys.}\ }\textbf {\bibinfo {volume} {139}},\ \bibinfo {pages} {021102} (\bibinfo {year} {2013})}\BibitemShut {NoStop}%
\bibitem [{\citenamefont {Kats}(2014)}]{kats_dcsd_2014}%
  \BibitemOpen
  \bibfield  {author} {\bibinfo {author} {\bibfnamefont {D.}~\bibnamefont {Kats}},\ }\bibfield  {title} {\enquote {\bibinfo {title} {Communication: The distinguishable cluster approximation. {II}. the role of orbital relaxation},}\ }\href {\doibase 10.1063/1.4892792} {\bibfield  {journal} {\bibinfo  {journal} {J. Chem. Phys.}\ }\textbf {\bibinfo {volume} {141}},\ \bibinfo {pages} {061101} (\bibinfo {year} {2014})}\BibitemShut {NoStop}%
\bibitem [{\citenamefont {McArdle}\ and\ \citenamefont {Tew}(2020)}]{McArdle2020}%
  \BibitemOpen
  \bibfield  {author} {\bibinfo {author} {\bibfnamefont {S.}~\bibnamefont {McArdle}}\ and\ \bibinfo {author} {\bibfnamefont {D.~P.}\ \bibnamefont {Tew}},\ }\href@noop {} {\enquote {\bibinfo {title} {Improving the accuracy of quantum computational chemistry using the transcorrelated method},}\ } (\bibinfo {year} {2020}),\ \bibinfo {note} {arXiv:2006.11181}\BibitemShut {NoStop}%
\bibitem [{\citenamefont {Sokolov}\ \emph {et~al.}(2023)\citenamefont {Sokolov}, \citenamefont {Dobrautz}, \citenamefont {Luo}, \citenamefont {Alavi},\ and\ \citenamefont {Tavernelli}}]{Sokolov2020}%
  \BibitemOpen
  \bibfield  {author} {\bibinfo {author} {\bibfnamefont {I.~O.}\ \bibnamefont {Sokolov}}, \bibinfo {author} {\bibfnamefont {W.}~\bibnamefont {Dobrautz}}, \bibinfo {author} {\bibfnamefont {H.}~\bibnamefont {Luo}}, \bibinfo {author} {\bibfnamefont {A.}~\bibnamefont {Alavi}}, \ and\ \bibinfo {author} {\bibfnamefont {I.}~\bibnamefont {Tavernelli}},\ }\bibfield  {title} {\enquote {\bibinfo {title} {Orders of magnitude increased accuracy for quantum many-body problems on quantum computers via an exact transcorrelated method},}\ }\href {\doibase 10.1103/PhysRevResearch.5.023174} {\bibfield  {journal} {\bibinfo  {journal} {Phys. Rev. Res.}\ }\textbf {\bibinfo {volume} {5}},\ \bibinfo {pages} {023174} (\bibinfo {year} {2023})}\BibitemShut {NoStop}%
\bibitem [{\citenamefont {Dobrautz}\ \emph {et~al.}(2024)\citenamefont {Dobrautz}, \citenamefont {Sokolov}, \citenamefont {Liao}, \citenamefont {Ríos}, \citenamefont {Rahm}, \citenamefont {Alavi},\ and\ \citenamefont {Tavernelli}}]{Dobrautz2024}%
  \BibitemOpen
  \bibfield  {author} {\bibinfo {author} {\bibfnamefont {W.}~\bibnamefont {Dobrautz}}, \bibinfo {author} {\bibfnamefont {I.~O.}\ \bibnamefont {Sokolov}}, \bibinfo {author} {\bibfnamefont {K.}~\bibnamefont {Liao}}, \bibinfo {author} {\bibfnamefont {P.~L.}\ \bibnamefont {Ríos}}, \bibinfo {author} {\bibfnamefont {M.}~\bibnamefont {Rahm}}, \bibinfo {author} {\bibfnamefont {A.}~\bibnamefont {Alavi}}, \ and\ \bibinfo {author} {\bibfnamefont {I.}~\bibnamefont {Tavernelli}},\ }\bibfield  {title} {\enquote {\bibinfo {title} {Toward real chemical accuracy on current quantum hardware through the transcorrelated method},}\ }\href {\doibase 10.1021/acs.jctc.4c00070} {\bibfield  {journal} {\bibinfo  {journal} {J. Chem. Theory Comput.}\ }\textbf {\bibinfo {volume} {20}},\ \bibinfo {pages} {4146--4160} (\bibinfo {year} {2024})}\BibitemShut {NoStop}%
\bibitem [{\citenamefont {Simula}\ \emph {et~al.}()\citenamefont {Simula}, \citenamefont {Christlmaier}, \citenamefont {Filip}, \citenamefont {Haupt}, \citenamefont {Kats}, \citenamefont {{López Ríos}},\ and\ \citenamefont {Alavi}}]{Simula2024}%
  \BibitemOpen
  \bibfield  {author} {\bibinfo {author} {\bibfnamefont {K.}~\bibnamefont {Simula}}, \bibinfo {author} {\bibfnamefont {E.~M.~C.}\ \bibnamefont {Christlmaier}}, \bibinfo {author} {\bibfnamefont {M.-A.}\ \bibnamefont {Filip}}, \bibinfo {author} {\bibfnamefont {J.~P.}\ \bibnamefont {Haupt}}, \bibinfo {author} {\bibfnamefont {D.}~\bibnamefont {Kats}}, \bibinfo {author} {\bibfnamefont {P.}~\bibnamefont {{López Ríos}}}, \ and\ \bibinfo {author} {\bibfnamefont {A.}~\bibnamefont {Alavi}},\ }\href@noop {} {\bibinfo  {journal} {in preparation}\ }\BibitemShut {NoStop}%
\end{thebibliography}%

\end{document}